\documentstyle[12pt,amssymb]{article}
\SetSymbolFont{largesymbols}{normal}{OMX}{lcmex}{m}{n}
  
\catcode`@=11
\def\secteqno{\@addtoreset{equation}{section}%
\def\theequation{\thesection.\arabic{equation}}}
\catcode`@=12
\secteqno
\def\dd{\hbox{\,\Large$\triangleright$}}
\newcommand{\be}{\begin{equation}}
\newcommand{\ee}{\end{equation}}
\newcommand{\bea}{\begin{eqnarray}}
\newcommand{\eea}{\end{eqnarray}}
\newcommand{\bref}[1]{(\ref{#1})}
\newcommand{\nn}{\nonumber}

\newcommand{\slP}{/ {\hskip-0.27cm{P}}}

\def\dig#1{\setbox0=\hbox{$#1M$}
	\hskip.06\wd0 \vrule width.07\wd0 height.63\wd0 depth.01\wd0 
	\vrule width.37\wd0 height.63\wd0 depth-.56\wd0 \hskip-.4\wd0
	\vrule width.25\wd0 height.35\wd0 depth-.28\wd0 
	\vrule width.07\wd0 height.35\wd0 depth-.17\wd0 \hskip.14\wd0}
\def\digamma{{\mathpalette\dig{}}}
\def\pa{\partial}\def\={{\;=\;}}\def\+{{\;+\;}}

\begin{document}

\begin{flushright}
\parbox{4.2cm}
{2015,~July 11
 \\
KEK-TH-1839 \hfill \\YITP-SB-15-21
\hfill \\
}
\end{flushright}

\vspace*{1.1cm}

\begin{center}
 {\Large\bf Type II chiral affine Lie algebras }\\
 {~}\\
 {\Large\bf  and string actions in doubled  space  }
\end{center}
\vspace*{1.5cm}
\centerline{\large Machiko Hatsuda$^{\dagger^\natural }$\footnote{mhatsuda@post.kek.jp
}, Kiyoshi Kamimura$^{\dagger }$\footnote{kamimura@ph.sci.toho-u.ac.jp}
and Warren Siegel$^\star$
\footnote{siegel@insti.physics.sunysb.edu
,
http://insti.physics.sunysb.edu/{\tt \~{}}siegel/plan.html}
}
\begin{center}
$^{\dagger}$\emph{Physics Division, Faculty of Medicine,
 Juntendo University, Chiba 270-1695, Japan}
\\
$^{\natural}$\emph{KEK Theory Center, High Energy Accelerator Research 
Organization,\\
Tsukuba, Ibaraki 305-0801, Japan} 
\\
$^\star$\emph{C. N. Yang Institute for Theoretical Physics
State University of New York, Stony Brook, NY 11794-3840}
\vspace*{0.5cm}
\\
\end{center}

\vspace*{1cm}

\centerline{\bf Abstract}
\vspace*{0.5cm}
We present affine Lie algebras generated by 
the supercovariant derivatives and the supersymmetry generators
 for the left and right moving modes in the doubled space.
 Chirality  is manifest in our doubled space  as well as the T-duality
 symmetry.
 We present  gauge invariant bosonic and superstring actions 
 preserving the two-dimensional diffeomorphism
invariance  and the $\kappa$-symmetry 
where doubled spacetime coordinates are chiral fields.  
The doubled space becomes the usual space
by dimensional reduction constraints.
\vfill
\thispagestyle{empty}
\setcounter{page}{0}
\newpage

\section{Introduction}

The low energy effective field theory of the string theory has the T-duality symmetry.
The worldsheet origin of the T-duality is mixing of momenta and winding modes of a string.
These string modes are generalized to the supercovariant derivatives
for a superstring  which satisfy the affine super-Lie algebra \cite{Siegel:85}-
\cite{Siegel:1994xr}. 
Doubling spacetime coordinates 
makes the T-duality manifest
\cite{Duff:1989tf}-
\cite{Siegel:1993xq}.
A manifestly T-duality formulation based on the affine Lie algebra
in the doubled space was proposed in \cite{Siegel:1993xq}.
The affine Lie algebra determines 
 the new Lie bracket  or equivalently
 C-bracket  which gives rise to the stringy modification of the 
general coordinate transformation.
The generalized geometry proposed  in \cite{Hitchin:2004ut} 
is described by the Courant bracket 
which is reduced  from the C-bracket by the dimensional reduction.
The generalized geometry and the double field theory 
have been widely studied  
\cite{Hull:2004in}-
\cite{Blumenhagen:2014gva}
  and review articles are \cite{Aldazabal:2013sca,Hohm:2013bwa}.

Recently we have proposed a manifestly T-duality formulation
for a type II superstring 
\cite{Hatsuda:2014qqa,Pola2014}
 and the one with the Ramond-Ramond gauge fields
\cite{Hatsuda:2014aza}.
For a type II supersymmetric extension 
of the manifestly T-duality formalism 
chiral separation of affine Lie algebras is an essential problem.
We specify chirality  as  left and right moving modes in
a two dimensional worldsheet.
Chiral currents  for a bosonic string in a non-abelian background
  are constructed  by 
the Wess-Zumion-Witten   model \cite{Witten:1983ar}.
A group element  $\underline{g}=
 g(\sigma^+)g'(\sigma^-)$ is considered 
where  
$\sigma^\pm$ are the left and right moving two-dimensional coordinates.
The left moving current is constructed as the right-invariant current,
 $\partial_+ \underline{g}\underline{g}^{-1}=\partial_+ {g}{g}^{-1}(\sigma^+)
 $, while the right moving current is  constructed as
the left-invariant current,
  $\underline{ g}^{-1}\partial_- \underline{g}= { g'}^{-1}\partial_- {g'}(\sigma^-)$.
 However  it is known that 
 the local  supersymmetry generator (supercovariant derivative) is obtained from the left-invariant current,
 while the global supersymmetry generator is obtained from
 the right-invariant current for supersymmetric theories.
 For a type II superstring theory 
both the supercovariant derivative and the supersymmetry
generator must have  both the left and right moving modes,
not one for each.
Although the chiral separation of the supercovariant derivative algebra
for a superstring on the anti-de Sitter space 
is obtained on the constrained surface \cite{Hatsuda:2001xf},
the chiral separation of currents for a superstring 
 in nonabelian backgrounds is in general  still difficult.  

 In our formulation 
 we begin by two independent Lie groups 
G and G' which are parameterized by $Z^M$ and $Z^{M'}$.
For a direct product of the groups G$\times$G'
a group element satisfies $\underline{g}=  g(Z^M)g'(Z^{M'}) =g'(Z^{M'}) g(Z^M) $.
Therefore the left-invariant and the right-invariant one forms 
contain both the left and right moving modes as
 $\underline{ g}^{-1}d \underline{g}=g^{-1}dg+g'{}^{-1}dg'=J(Z^M)+J(Z^{M'}) $ and 
 $d\underline{g}\underline{g}^{-1}
 =dgg^{-1}+dg'g'{}^{-1}=\tilde{J}(Z^M)+\tilde{J}(Z^{M'})$.
This is similar to the nonabelian currents given by Tseytlin \cite{Tseytlin:1990nb}.
The chiral scalar action used in that paper 
does not preserve the two-dimensional diffeomorphism invariance  \cite{Siegel:1983es}.
In our formulation a chiral scalar action preserves 
 the two-dimensional  diffeomorphism invariance allowing the $\kappa$-symmetry
 for the superstring action.
The doubled coordinates are chiral fields, $Z^M(\sigma^+)$ and $Z^{M'}(\sigma^-)$.
The stringy geometry is governed by the affine Lie algebras
generated by the chiral supercovariant derivatives.
The covariant derivatives are still
manifestly chiral even after reducing into the usual space
by dimensional reduction constraints, since the dimensional reduction constraints
are given by the auxiliary symmetry generators.

In order to construct a doubled space 
 there are two types of doubling of a group G:
\begin{enumerate}
  \item{Semidirect product, G$\to$ G$\ltimes$G$^\ast$:

A Lie group G is generated  by ${\mathfrak g}$ corresponding to 
derivative operators which include  momenta,
while another group  G$^\ast$ is generated by ${\mathfrak g}^\ast$ 
corresponding to one form currents 
which include winding modes.
This is a conventional way of doubling  
discussed in for example \cite{Cremmer:1998px,Hitchin:2004ut,Ruiz:2014fta}. 
It gives the following inhomogeneous algebra
\bea
\lbrack{\mathfrak g} ,{\mathfrak g}]={\mathfrak g}~~,~~[{\mathfrak g},{\mathfrak g}^\ast]={\mathfrak g}^\ast~~,~~[{\mathfrak g}^\ast,{\mathfrak g}^\ast]=0~~~.\label{GGast}
\eea }
\item{Direct product,  G$\to$G$\times$G': 

 G and G' are independent copy and they are
 generated by  ${\mathfrak g}$ and ${\mathfrak g}'$ corresponding to the
 left and right moving modes respectively. 
 They satisfy the following algebra 
\bea
\lbrack
{\mathfrak g},{\mathfrak g}]={\mathfrak g}~~,~~[{\mathfrak g},{\mathfrak g}']=0~~,~~[{\mathfrak g}',{\mathfrak g}']={\mathfrak g}'~~.\label{GGpr}
\eea
It is straightforward to construct one form currents and derivative operators
which satisfy \bref{GGast}.
In terms of  them 
we present a general construction of chiral currents satisfying 
\bref{GGpr}. This construction requires that 
the Lie group must have a nondegenerate group metric and 
the Lie algebra must have grading by the canonical dimension. 
}
\end{enumerate}

The organization of the paper is the following:
In the next section we present 
a general construction of chiral affine Lie algebras
generated  by the supercovariant derivative and the symmetry generator.
We begin by a Lie algebra with
a nondegenerate group metric \cite{Polacek:2013nla,Bonanos:2009wy}. 
It is necessary to include the Lorentz $S_{mn}$ and 
its nondegenerate partner $\Sigma^{mn}$
for construction of two independent affine Lie algebras 
generated by the covariant derivative, $P_m$, 
and the symmetry generator, $\tilde{P}_m$. 
The Lie algebra is graded by a dilatation operator 
which plays an important role in this construction.
Generators of the chiral affine algebras 
include the $B$ field.
For a flat background the
$B$ field can be written in terms of the dilatation operator.

In section 3 
we present  chiral affine Poincar{\'e} algebras in the doubled space and 
concrete expression of generators.
A set of dimensional reduction constraints are examined 
which reduce the doubled space into the usual space with preserving 
the local geometry.
We present  an gauge invariant action for a bosonic string in the doubled  space.
Then we demonstrate that the string action in the doubled space 
is reduced into the usual action.
In section 4 we present chiral affine super-Poincar{\'e} algebras in the doubled space.
Then an gauge invariant action for a type II superstring in the doubled space is given.
The super-doubled space is  also reduced into the usual  space.


\section{Chiral affine Lie algebras}

In this section we present a general construction of 
two sets of affine Lie algebras generated by the
covariant derivatives and the symmetry generators.
\begin{itemize}
   \item{Affine Lie algebras
    
  \begin{tabular}{l}
  \\\\
{Lie~algebra}~$G_I$
\end{tabular}
\begin{tabular}{clccc}
&&particle&$\to$&
string\\&&&&~affine Lie algebra
\\
$\nearrow$&{covariant~derivative}&$\mathring{\nabla}_I$&$\to$&$\mathring{\dd}_I(\sigma)$\\
$ \searrow$ &{symmetry~generator}&$\tilde{\nabla}_I$&$\to$&$\tilde{\dd}_I(\sigma)$\\
\end{tabular}
}
\end{itemize}
Next we double the Lie algebra as
a direct product:
\begin{itemize}
 
\item{Doubled chiral affine Lie algebras

  \begin{tabular}{l}
  \\
{Lie~algebra}~$G_I$~$\to$~
$G_M,~G_{M'}$~ 
\end{tabular}
\begin{tabular}{clcc}
&&left&
right~
\\
$\nearrow$&{covariant~derivative}&$\mathring{\dd}_M(\sigma^+)$&
$\mathring{\dd}_{M'}(\sigma^-)$\\
$ \searrow$ &{symmetry~generator}&$\tilde{\dd}_M(\sigma^+)$&
$\tilde{\dd}_{M'}(\sigma^-)$\\
\end{tabular}

 }
\end{itemize} 
The doubled coordinates manifest chirality as well as the T-duality symmetry.

In subsection 2.1 at first  for a given Lie algebra we construct 
the left-invariant current $J$, the right-invariant current $\tilde{J}$,   
the particle covariant derivative $\mathring{\nabla}$, and the particle
symmetry generator $\tilde{\nabla}$.
The derivatives and the currents satisfy the case 1 algebra
in \bref{GGast}. 
The canonical dimensions of operators  
are expressed by an eigenvalue matrix of the dilatation operator. 
In subsection 2.2 the general construction of 
 affine Lie algebras for the string covariant derivative $\mathring{\dd}$,
 and the string symmetry generator $\tilde{\dd}$.
 They are linear combinations of the particle derivatives, $\mathring{\nabla}$, $\tilde{\nabla}$,
 and the $\sigma$-components of the currents, $J_1,\tilde{J}_1$,  
 with the $B$ field as  coefficients.
 In subsection 2.3 the Lie algebra is doubled.
This doubling corresponds to 
the case 2 algebra in \bref{GGpr}.
The way of the doubling gives 
chiral affine Lie algebras.

\subsection{Derivative operators and one form currents}

A Lie algebra is  generated  by 
 $G_I$ with
  \bea
&\lbrack G_I,G_J\}=if_{IJ}{}^KG_K&\label{alg}~~~,
 \eea
 where $[A,B\}=AB-(-)^{AB}BA$ is the graded commutator.
A nondegenerate group metric $\eta_{IJ}$ is introduced 
in such a way that the structure constant $f_{IJ}{}^K$ with lowered indices
becomes totally graded antisymmetric
\bea
&{\rm tr}(G_IG_J)=\eta_{IJ}=\displaystyle\frac{1}{2}\eta_{(IJ]}~~,~~
f_{IJK}\equiv f_{IJ}{}^L\eta_{LK}=
\displaystyle\frac{1}{3!}f_{[IJK)}&~~~.\label{GGeta}
\eea
We introduce a dilatation operator $\hat{N}$ 
whose eigenvalues are canonical dimensions $n_I$ as 
\bea
\lbrack \hat{N},G_I]=iN_{I}{}^JG_J=in_IG_I\label{candim}~~.
\eea
The Jacobi identity of the dilatation operator, $\hat{N}$, and Lie algebra generators,
$G_I$'s , 
gives an identity
\bea
&&\lbrack \hat{N},[G_I,G_J\}]+
 \lbrack G_I,[G_J,\hat{N}]\}+
 (-)^{IJ} \lbrack G_J,[ \hat{N},G_I]\}=0\nn\\
&&~\Rightarrow~
-N_{[J}{}^Kf_{I)KL}+f_{IJ}{}^{K}N_K{}^M\eta_{ML}=(\tilde{n}_L-n_I-n_J)
f_{IJL}=0\label{dilajacobi}~~
\eea
with
\bea
\tilde{n}_I~G_I\equiv \eta^{LJ} N_J{}^K \eta_{KI}G_L
~~.
\eea
The sum of  canonical dimensions of a nondegenerate pair  is set to be $n_0$,
and the sum of  canonical dimensions of the lower indices of the structure constant
in \bref{dilajacobi} becomes also $n_0$ 
\bea
&&(N\eta +\eta N^T)_{IJ}=(n_I+n_J)\eta_{IJ}=n_0 \eta_{IJ}\nn\\
&&\Rightarrow 
\tilde{n}_I=(n_0-n_I)~~,~~
(n_0-{n}_K-n_I-n_J)
f_{IJK}=0~~~.
\eea
We take $n_0=2$ in order to choose $n_P=1$ for $\eta_{PP}=1$.

Particle derivatives of the Lie algebra $\mathring{\nabla}_I$ and $\tilde{\nabla}_I$ 
constructed at first. 
An element of the Lie group $g=g(Z)$ is parameterized by the coordinates $Z^M$.
There are two kinds of currents which are invariant under
right or left multiplicative actions:
\begin{itemize}
  \item {The left-invariant current and 
  the particle covariant derivative
  generating the right multiplication $g_0\to g_0 g$ :
  \bea
&&  J=g^{-1}dg
   =iJ^{{I}}G_{{I}}~~,~~
  dJ=-J\wedge J~~,~~
  J^{{I}}  =dZ^{{M}}R_{{M}}{}^{{I}}
  \nn\label{rc}~\\
&& \mathring{\nabla}_{{I}}=(R^{-1})_{{I}}{}^{{M}}\frac{1}{i}\partial_{{M}} ~~,~~
  ~[\mathring{\nabla}_I,\mathring{\nabla}_J\}
  =-if_{IJ}{}^K\mathring{\nabla}_K ~~~\nn\\\label{Cvd}
  &&~~\Rightarrow~~
\partial_{[M}R_{N)}{}^K=(-)^{NI}\frac{1}{2}R_{[M}{}^IR_{N)}{}^Jf_{JI}{}^K~
  \eea
  }
  \item{The right-invariant current (Noether current) 
  and the particle symmetry generator
  generating the left multiplication $g_0\to gg_0$
:
  \bea
 && \tilde{J}=(dg)g^{-1}=i\tilde{J}^{{I}}G_{{I}}~~,~~  
 d\tilde{J}=\tilde{J}\wedge \tilde{J}~~\nn
,~~\tilde{J}^{{I}}=dZ^{{M}}L_{{M}}{}^{{I}}~~\nn\\
&&\tilde{\nabla}_{{I}}=(L^{-1})_{{I}}{}^{{M}}\frac{1}{i}\partial_{{M}} 
 ~~,~~\lbrack \tilde{\nabla}_I,\tilde{\nabla}_J\}
=if_{IJ}{}^K\tilde{\nabla}_K
 \nn\\&&~~\Rightarrow~~
\partial_{[M}L_{N)}{}^K=-(-)^{NI}\frac{1}{2}L_{[M}{}^IL_{N)}{}^Jf_{JI}{}^K
 .\label{lc}
  \eea
  }
 \item{Independence of the right and left multiplications:
 \bea
& & [\mathring{\nabla}_I,\tilde{\nabla}_J\}
  =0 ~~\to~~(\partial_MR_N{}^I)R^{-1}{}_I{}^L
=(-)^{NM}(\partial_NL_M{}^I)L^{-1}{}_I{}^L
  ~~~.\label{Cvd2}
 \eea}
\end{itemize}

Particle derivatives are extended to the string affine algebra generators;
$\mathring{\nabla}_I\to \mathring{\nabla}_I(\sigma)$ and
$\tilde{\nabla}_I\to \tilde{\nabla}_I(\sigma)$.
$\tau$ and $\sigma$ components of currents are denoted by
$J^I=d\sigma^iJ_i^I=d\tau J_0^I+d\sigma J_1^I$ and 
$\tilde{J}^I=d\tau \tilde{J}_0^I+d\sigma \tilde{J}_1^I$.
Currents $J_1^I$ and $\tilde{J}_1^I$ carry the canonical dimension
$2-n_I$, since $\partial_\sigma$ carries the canonical dimension 2 
where $\alpha'$ is abbreviated.
The indices of currents are lowered with $\eta_{IJ}$ as 
$J_{{I}}\equiv J^{{L}}\eta_{{LI}}$ and 
$\tilde{J}_{{I}}\equiv \tilde{J}^{{L}}\eta_{{LI}}$,
and they are covariant under 
$\mathring{\nabla}_I$ and $\tilde{\nabla}_I$
respectively.
Derivatives and currents satisfy 
the case 1. semidirect product   G$\ltimes$G$^\ast$ in
 \bref{GGast}, where
$ \mathring{\nabla}\in {\mathfrak g}$, $J_1^I\in{\mathfrak g}^\ast$ 
and  $\tilde{\nabla}\in {\mathfrak g}$, $\tilde{J}_1^I\in{\mathfrak g}^\ast$; 
\bea
&&{\renewcommand{\arraystretch}{1.6}
\left\{\begin{array}{lcl}
  [\mathring{\nabla}_I(1),\mathring{\nabla}_J(2)\}
  &=&-if_{IJ}{}^K\mathring{\nabla}_K\delta(2-1)
\label{pcov}\\
\lbrack \mathring{\nabla}_{ {I}}(1),J_1^{ {J}}(2)\}&=&
-if_{KI}{}^J J^K_1\delta(2-1)
-i\delta_I^J\partial_\sigma\delta(2-1)\\
\lbrack J_1^I(1),J_1^J(2)\}&=&0\end{array}\right.}\\
&&{\renewcommand{\arraystretch}{1.6}
\left\{
\begin{array}{lcl}
\lbrack \tilde{\nabla}_I(1),\tilde{\nabla}_J(2)\}
&=&if_{IJ}{}^K\tilde{\nabla}_K\delta(2-1)~~~
\label{psym}\\
\lbrack \tilde{\nabla}_{ {I}}(1),\tilde{J}_1^{ {J}}(2)\}&=&
if_{KI}{}^J \tilde{J}_1^{K}\delta(2-1)
+i\delta_I^J\partial_\sigma\delta(2-1)
\\  
\lbrack \tilde{J}_1^I(1),\tilde{J}_1^J(2)\}&=&0\end{array}\right.}\\
&&{\renewcommand{\arraystretch}{1.6}\left\{
\begin{array}{lcl}
\lbrack \mathring{\nabla}_{{I}}(1),\tilde{\nabla}_{J}(2)\}&=&0\\
\lbrack \tilde{\nabla}_{I}(1),{J}_1^{J}(2)\}&=&
-iM_I{}^J(2)\partial_\sigma\delta(2-1)\\
\lbrack \mathring{\nabla}_{I}(1),\tilde{J}_1^{J}(2)\}&=&
-i(M^{-1}){}_I{}^J(2)\partial_\sigma\delta(2-1)\end{array}\right.
 ~~~\label{pcovsym}}
\eea
The $\sigma$ coordinates of the two-dimensional space, 
$\sigma_1$ and $\sigma_2$, 
are denoted by $1$ and $2$.
The commutator between
$\mathring{\nabla}_I$ and $\tilde{J}^I$ and the one between   
 $\tilde{\nabla}_I$ and ${J}^I$ include a matrix $M_I{}^J$
 which is defined as
\bea
M_I{}^J=(L^{-1})_I{}^MR_M{}^J
~~~.
\eea
It relates the left-invariant and the right-invariant currents, and
it also relates  
the covariant derivative and the symmetry generator as
\bea
\tilde{J}^IM_I{}^K=J^K~~,~~
\tilde{\nabla}_I=M_I{}^K\mathring{\nabla}_K~~
\label{gJg}.
\eea 
From the relation between currents $\tilde{J}$ and $J $ in 
 \bref{rc}
 and \bref{lc} 
\bea
&g^{-1}\tilde{J}g=g^{-1}(idZ^ML_M{}^IG_I)g=idZ^MR_M{}^IG_I=J
\Rightarrow~~
~g^{-1}~G_I~g=M_I{}^JG_J
\label{ML1R}~~,\nn&\eea
$M_I{}^J$ satisfies an orthonormal condition  as
\bea
&\eta_{IJ}={\rm tr}(G_IG_J)={\rm tr}(g^{-1}G_Igg^{-1}G_Jg)=M_I{}^LM_{J}{}^{K}
{\rm tr}(G_LG_K)&\nn\\
&\Rightarrow
~~\eta_{IJ}=M_I{}^L(-)^{K(L+K)}M_{J}{}^{K}\eta_{LK}
=M_I{}^L(M^T)^{K}{}_{J}\eta_{LK}
 ~~~.&\label{MMTeta}
\eea
The super-transpose of a matrix is given as
$(M^T)^K{}_J=(-)^{K(J+K)}M_{J}{}^{K}$.
From now on the super-transpose  is implicitly imposed. 
\bref{gJg} together with \bref{MMTeta} leads to 
\bea
M_I{}^{P}M_J{}^{Q}M_K{}^Rf_{PQR}=f_{IJK}~~,~~J^IJ^J\eta_{IJ}=\tilde{J}^I\tilde{J}^J\eta_{IJ}.
\label{RLRL}
\eea 
Relations \bref{rc}, \bref{lc} and \bref{Cvd2} lead to that 
 $M_{I}{}^J$ is a function of $Z^M$ satisfying
 \bea
 i\tilde{\nabla}_K 
 M_I{}^J=f_{IK}{}^LM_L{}^J~~,~~
 i\mathring{\nabla}_K
 M_I{}^J=M_I{}^Lf_{LK}{}^J~~~.\label{Mf}
 \eea

 \par
\vskip 6mm

\subsection{Affine Lie algebras and $B$ field}

We construct two independent sets of affine Lie algebras 
generated  by the covariant derivative and 
by the symmetry generator.
\begin{itemize}
  \item { Covariant derivative $\mathring{\dd}_{{I}}$:
  \bea
  \lbrack \mathring{\dd}_{{I}}(1), 
\mathring{\dd}_{{J}}(2)\}=-if_{{IJ}}{}^{{K}}
\mathring{\dd}_{{K}}\delta(2-1)-i\eta_{{IJ}}
\partial_\sigma\delta(2-1)\label{cov}
  \eea
}
  \item{{ Symmetry}
{generator} $\tilde{\dd}_{{I}}$:
\bea
\lbrack \tilde{\dd}_{{I}}(1), 
\tilde{\dd}_{{J}}(2)\}=if_{{IJ}}{}^{{K}}
\tilde{\dd}_{{K}}\delta(2-1)+i\eta_{{IJ}}
\partial_\sigma\delta(2-1)\label{sym}
\eea
}
\item{
The covariant derivative $\mathring{\dd}_{{I}}$
 and the symmetry generator $\tilde{\dd}_{{I}}$ commute
with each other,
  \bea
  \lbrack \mathring{\dd}_{{I}}(1), 
\tilde{\dd}_{{J}}(2)\}=0~~~.\label{covsymzero}
  \eea
}
\end{itemize}
Jacobi identities of these affine algebras \bref{cov} and \bref{sym} 
require the existence of the nondegenerate group metric $\eta_{IJ}$
and the totally antisymmetricity of the structure constant $f_{IJK}$ in \bref{GGeta}.

Let us set the generators of the affine algebras  \bref{cov} and \bref{sym}
in terms of  currents and derivatives  in \bref{rc}, \bref{lc}, 
\bref{pcov} and \bref{psym} as
\bea
&\mathring{\dd}_{I}=\mathring{\nabla}_{I}
+J_1^{K}b_{KI}~~,~~
\tilde{\dd}_I=\tilde{\nabla}_{I}+\tilde{J}_1^K\tilde{b}_{KI}~~~.&\label{ddJb}
\eea
The algebras  \bref{cov}, \bref{sym} and \bref{covsymzero} give
conditions on $b_{IJ}$ and $\tilde{b}_{IJ}$:
The symmetric parts of $b_{IJ}$ and $\tilde{b}_{IJ}$
are uniquely determined from the signature of the Schwinger terms,
the term including
 $\partial_\sigma\delta(2-1)$, 
of the affine Lie algebras \bref{cov} and \bref{sym},
\bea
b_{(IJ]}=\eta_{IJ}=-\tilde{b}_{(IJ]}~~~.\label{bIJ}
\eea
Vanishing the coefficient of $\partial_\sigma\delta(2-1)$ in  \bref{covsymzero}
leads to
\bea
\tilde{b}_{KI}=-b_{JL}M_I{}^JM_K{}^L~~~.\label{btilb}
\eea
Vanishing the coefficient of $\delta(2-1)$ in  \bref{covsymzero}
using with \bref{RLRL} leads to
\bea
i\mathring{\nabla}_{[J}
 b_{I)K}-
 i\mathring{\nabla}_{K}
 b_{IJ}
-\eta^{LM}b_{M[J}f_{I)LK}+f_{IJ}{}^Lb_{KL}
&=&0\label{bIJIJ}~~\\
i\tilde{\nabla}_{[J|}
 \tilde{b}_{K|I)}-
 i\tilde{\nabla}_{K}
 \tilde{b}_{JI}
-\eta^{ML}\tilde{b}_{[J|L}f_{|I)MK}+f_{IJ}{}^L\tilde{b}_{LK}
&=&0~~\nn
~.
\eea
A simple solution of $b_{IJ}$ and $\tilde{b}_{IJ}$ is
obtained from the Jacobi relation for $N_I{}^J$ given in \bref{dilajacobi}
as 
 \bea
b_{KI}&=&\frac{1}{2}N_I{}^L\eta_{LK}=\frac{1}{2}
(\eta_{KI}+B_{KI})~~\nn\\
B_{IK}&=&-(-)^{IK}B_{KI}=\frac{1}{2}N_{[K|}{}^L\eta_{L|I)}
=\frac{1}{2}(n_K-n_I)\eta_{IK}\label{BnB}\\
\tilde{b}_{KI}&=&-\frac{1}{2}N_L{}^M\eta_{MJ}M_I{}^JM_K{}^L
=-\frac{1}{2}(\eta_{KI}
+M_K{}^LM_I{}^JB_{JL})~~~.\nn
\eea
In this solution $B_{IJ}$ is constant and $M_{I}{}^J$ depends on 
parameters.
There is 
 ambiguity in solutions of $b_{IJ}$ and $\tilde{b}_{IJ}$,
which can be interchanged.
The generators of the affine algebras are given as the followings:
\bea
{\renewcommand{\arraystretch}{1.6}
\begin{array}{ccl}
\mathring{\dd}_{I}&=&\mathring{\nabla}_{I}
+\frac{1}{2}J_1^{K}N_I{}^L\eta_{LK}
=\mathring{\nabla}_{I}
+\frac{1}{2}(J_1^{K}\eta_{KI}+J_1^KB_{KI})
~\label{symbb}\\
\tilde{\dd}_I&=&\tilde{\nabla}_{I}
-\frac{1}{2}\tilde{J}_1^K
 M_{K}{}^{L}N_{L}{}^M\eta_{MJ}M_I{}^{J}
= \tilde{\nabla}_{I}
+\frac{1}{2}(-\tilde{J}_1^K\eta_{KI}
+J_1{}^KB_{KJ}M_I{}^J)
\end{array}}
\label{covsymbb}
\eea

The $B$ field in the Wess-Zumino term is written as 
\bea
B&=&\frac{1}{2}dZ^M\wedge dZ^NB_{MN}=J^I\wedge J^Jb_{IJ}
=\tilde{J}^I\wedge \tilde{J}^J\tilde{b}_{IJ}~\\
B_{MN}&=&\frac{1}{2}R_M{}^IR_N{}^Jb_{[IJ)}=
\frac{1}{2} L_M{}^IL_N{}^J\tilde{b}_{[IJ)}~~.\nn
\eea
The three form $H=dB$ is written  from \bref{bIJIJ} as
\bea
H=dB=\frac{1}{3!}f_{IJK}J^I\wedge J^J\wedge J^K
=\frac{1}{3!}f_{IJK}\tilde{J}^I\wedge \tilde{J}^J\wedge \tilde{J}^K~~~,
\eea
and it is closed 
\bea
dH=\frac{1}{2}f_{IJK}dJ^I\wedge J^J\wedge J^K=\frac{1}{2}f_{IJK}f^I{}_{LM}J^L\wedge J^M
\wedge J^J\wedge J^K=0~~~.
\eea
\par
\vskip 6mm
\subsection{``Chirality" from doubling}

Now we double the algebra  \bref{alg} together with 
the nondegenerate metric \bref{GGeta}
in the second way of doubling in \bref{GGpr} G$\times$G'.
Indices are denoted by $_{\underline{M}}=_{(M,M')}$ corresponding to
the left and right sectors.
The structure constant $f_{\underline{MN}}{}^{\underline{L}}$, the nondegenerate group metric,
$\eta_{\underline{MN}}$ are extended as follows. 
Another metric  
 $\hat{\eta}_{\underline{MN}}$ is  introduced to define the Virasoro operator. 
 \bea
{\renewcommand{\arraystretch}{1.6}
\begin{array}{ccl}
{\rm G}&\to&{\rm G}\times {\rm G'}\\
_{I}&\to&_{\underline{M}}=_{(M,M')}\\
f_{IJ}{}^K&\to& f_{\underline{MN}}{}^{\underline{L}}=
(f_{MN}{}^L,~f_{M'N'}{}^{L'}=-f_{MN}{}^L)\\
\eta_{IJ}&\to&\eta_{\underline{M}\underline{N}}= (\eta_{MN},~\eta_{M'N'}=-\eta_{MN})~~~\\
\hat{\eta}_{IJ}&\to&\hat{\eta}_{\underline{M}\underline{N}}= (\hat{\eta}_{MN}=\eta_{MN},~\hat{\eta}_{M'N'}=\eta_{MN})
\end{array}}~~\label{etaetahat}
\eea
Then we double the affine Lie algebras generated by the covariant derivative  \bref{cov}
and the one by the symmetry generator \bref{sym}.
Doubled affine Lie algebras 
are given as below:
\begin{itemize}
  \item { Covariant derivative $\mathring{\dd}_{\underline{M}}$:
  \bea
  \lbrack \mathring{\dd}_{\underline{M}}(1), 
\mathring{\dd}_{\underline{N}}(2)\}&=&-if_{\underline{MN}}{}^{\underline{L}}
\mathring{\dd}_{\underline{L}}\delta(2-1)-i\eta_{\underline{MN}}
\partial_\sigma\delta(2-1)\label{covlr}
\eea
 \bea
{\renewcommand{\arraystretch}{1.6}
 \left\{\begin{array}{ccl}
  \lbrack \mathring{\dd}_{{M}}(1), 
\mathring{\dd}_{{N}}(2)\}&=&-if_{{MN}}{}^{{L}}
\mathring{\dd}_{{L}}\delta(2-1)-i\eta_{{MN}}
\partial_\sigma\delta(2-1)\label{cov2}\\
  \lbrack \mathring{\dd}_{{M'}}(1), 
\mathring{\dd}_{{N'}}(2)\}&=&if_{{MN}}{}^{{L}}
\mathring{\dd}_{{L'}}\delta(2-1)+i\eta_{{MN}}
\partial_\sigma\delta(2-1)\\
  \lbrack \mathring{\dd}_{{M}}(1), 
\mathring{\dd}_{{N}'}(2)\}&=&0~~~
\end{array}\right.}\nn
  \eea
}
  \item{{ Symmetry}
{generator} $\tilde{\dd}_{\underline{M}}$:
\bea
 \lbrack \tilde{\dd}_{\underline{M}}(1), 
\tilde{\dd}_{\underline{N}}(2)\}&=&if_{\underline{MN}}{}^{\underline{L}}
\tilde{\dd}_{\underline{L}}\delta(2-1)+i\eta_{\underline{MN}}
\partial_\sigma\delta(2-1)\label{sym2lr}
\eea
\bea
{\renewcommand{\arraystretch}{1.6}
 \left\{\begin{array}{ccl}
 \lbrack \tilde{\dd}_{{M}}(1), 
\tilde{\dd}_{{N}}(2)\}&=&if_{{MN}}{}^{{L}}
\tilde{\dd}_{{L}}\delta(2-1)+i\eta_{{MN}}
\partial_\sigma\delta(2-1)\label{sym2}\\
\lbrack \tilde{\dd}_{{M'}}(1), 
\tilde{\dd}_{{N'}}(2)\}&=&-if_{{MN}}{}^{{L}}
\tilde{\dd}_{{L'}}\delta(2-1)-i\eta_{{MN}}
\partial_\sigma\delta(2-1)\nn\\
 \lbrack \tilde{\dd}_{{M}}(1), 
\tilde{\dd}_{{N}'}(2)\}&=&0\end{array}\right.}
\eea
}
\item{
The covariant derivative $\mathring{\dd}_{{I}}$
 and the symmetry generator $\tilde{\dd}_{{I}}$ commute
with each other,
  \bea
  \lbrack \mathring{\dd}_{\underline{M}}(1), 
\tilde{\dd}_{\underline{N}}(2)\}&=&0~~~\label{covsymzero2}
  \eea
}
\end{itemize}
The signature of the Schwinger term $\partial_\sigma\delta(2-1)$ in \bref{covlr}
corresponds to the left or right chirality.

Two-dimensional diffeomorphisms are generated by the Virasoro operators,
${\cal H}_\tau$ and ${\cal H}_\sigma$.
The Virasoro operators and the Virasoro algebras  are given by 
\bea
&&{\renewcommand{\arraystretch}{1.6}
\left\{\begin{array}{cl}
{\cal H}_\tau&=\frac{1}{2}\mathring{\dd}_{\underline{M}}\hat{\eta}^{\underline{MN}}\mathring{\dd}_{\underline{N}}
\\
{\cal H}_\sigma&=\frac{1}{2}\mathring{\dd}_{\underline{M}}\eta^{\underline{MN}}\mathring{\dd}_{\underline{N}}
\end{array}\right.}
\\
&&{\renewcommand{\arraystretch}{1.6}
\begin{array}{ccl}
\lbrack {\cal H}_\tau(1),{\cal H}_\tau(2)]&=&
i({\cal H}_\sigma(1)+{\cal H}_\sigma(2))\partial_\sigma\delta(2-1)
\\
\lbrack {\cal H}_\sigma(1),{\cal H}_\tau(2)]&=&
i({\cal H}_\tau(1)+{\cal H}_\tau(2))\partial_\sigma\delta(2-1)
\\
\lbrack {\cal H}_\sigma(1),{\cal H}_\sigma(2)]&=&i({\cal H}_\sigma(1)
+{\cal H}_\sigma(2))\partial_\sigma\delta(2-1)~~~.
\end{array}}\label{ViraVira}
\eea
The two-dimensional  diffeomorphisms 
of a function of double coordinates $\Phi(Z^{\underline{M}})$ 
are given by 
\bea
&\partial_\tau \Phi=
i\lbrack \textstyle{\int}d\sigma~{\cal H}_\tau,\Phi]
~~,~~\partial_\sigma \Phi=
i\lbrack \textstyle{\int}d\sigma~{\cal H}_\sigma,\Phi]~~~.
&
\eea
The two dimensional left and right derivatives  
are given by
\bea
&&\partial_\pm=\frac{1}{2}(\partial_\tau\pm \partial_\sigma)\nn\\
&&\label{HH}
\Rightarrow~~
\partial_+\Phi=
i\lbrack \textstyle{\int}d\sigma~{\cal A},\Phi]
~~,~~\partial_- \Phi=
i\lbrack \textstyle{\int}d\sigma~{\cal A}',\Phi]
\\
&&
\left\{{\renewcommand{\arraystretch}{1.6}
\begin{array}{cll}
{\cal A}&=\frac{1}{2}({\cal H}_\tau+{\cal H}_\sigma)&=\frac{1}{2}\mathring{\dd}_{{M}}{\eta}^{{MN}}\mathring{\dd}_{{N}}
\\
{\cal A}'&=\frac{1}{2}({\cal H}_\tau-{\cal H}_\sigma)
&=\frac{1}{2}\mathring{\dd}_{{M}'}{\eta}^{{M'N'}}\mathring{\dd}_{{N'}}
\end{array}}
\right.~\nn\\
&&{\renewcommand{\arraystretch}{1.6}
\begin{array}{ccl}
\lbrack{\cal A}(1),{\cal A}(2)]&=&i({\cal A}(1)+{\cal A}(2))\partial_\sigma\delta(2-1)\\
\lbrack{\cal A}'(1),{\cal A}'(2)]&=&-i({\cal A}'(1)+{\cal A}'(2))\partial_\sigma\delta(2-1)
\label{AA}
\\
\lbrack{\cal A}(1),{\cal A}'(2)]&=&0
\end{array}}
\eea
Therefore the covariant derivatives satisfy the left or right chiral
conditions as 
\bea
{\renewcommand{\arraystretch}{1.6}
\begin{array}{ccl}
\partial_- \Phi(Z^M)=0&:&\Phi(Z^M)=(\mathring{\nabla}_M,~J^M,~\mathring{\dd}_M;
\tilde{\nabla}_M,~\tilde{J}^M,~\tilde{\dd}_M)\\
\partial_+ \Phi(Z^{M'})=0&:&\Phi(Z^{M'})=(\mathring{\nabla}_{M'},~J^{M'},~\mathring{\dd}_{M'};
\tilde{\nabla}_{M'},~\tilde{J}^{M'},~\tilde{\dd}_{M'})
\end{array}}
\label{tsA}
\eea
which comes from chiral property of the doubled coordinates $Z^{\underline{M}}$
\bea
\partial_- Z^M=0=\partial_+ Z^{M'}~\Rightarrow~
Z^M=Z^M(\sigma^+)~,~Z^{M'}=Z^{M'}(\sigma^-)~~~.\label{chiralZM}
\eea
For example the left-inariant current $J_i^M(\sigma^+) $ are functions of only $\sigma^+$
satisfying the Maurer-Cartan equation as 
\bea
\partial_+J_-^M-\partial_-J_+^M=J_+^NJ_-^Kf_{NK}{}^M~~\Rightarrow~~
\partial_-J_+^M(\sigma^+)=0~{\rm and}~J^M_-(\sigma^+)=0~~~.
\eea
Similar relations hold for other currents.

Let us consider the global  O(n,n) transformation 
where  the number of the dimensions
of the doubled space is 2n ($_{\underline{M}}$ runs 1 to 2n).
For a bosonic case n is given as
sum of dimensions of $P_m$, $S_{mn}$ and $\Sigma^{mn}$, 
 n=d+2$\times$$\frac{\rm d(d-1)}{2}$=d$^2$
where d is the number of the momenta.
Under the global O(n,n)$\ni {\cal O}_{\underline{M}}{}^{\underline{L}}$ the covariant derivative \bref{symbb} and \bref{etaetahat}
is transformed as
\bea
\mathring{\dd}_{\underline{M}}\to{\cal O}_{\underline{M}}{}^{\underline{L}}
\mathring{\dd}_{\underline{L}}~~,~~
{\cal O}_{\underline{M}}{}^{\underline{L}}\eta_{\underline{LK}}
({\cal O}^T)^{\underline{K}}{}_{\underline{N}}=\eta_{\underline{MN}}~~~.
\eea
Its affine Lie algebra \bref{covlr} is transformed covariantly preserving 
the structure 
of G$\times$G' as
\bea
~f_{\underline{MN}}{}^{\underline{L}}\to
{\cal O}_{\underline{M}}{}^{\underline{I}}{\cal O}_{\underline{M}}{}^{\underline{J}}f_{\underline{IJ}}{}^{\underline{K}}
({\cal O}^{-1})_{\underline{K}}{}^{\underline{L}}
~~~.
\eea
Under the global O(n,n) transformation the left and right 
moving modes are mixed and they are no  more chiral operators in general.
Under the O(n,n) transformation the two-dimensional $\sigma$-diffeomorphism constraint,
${\cal H}_\sigma$,
is inert, but the $\tau$-diffeoporphism constraint, ${\cal H}_\tau$,
causes the O(n,n) transformation on the gravitational background fields
$E_{\underline{A}}{}^{\underline{M}}$ as
\bea
{\cal M}^{\underline{MN}}=(E^T)^{\underline{M}}{}_{\underline{A}}\hat{\eta}^{\underline{AB}}
E_{\underline{B}}{}^{\underline{N}}
&\to&({\cal O}^T)^{\underline{M}}{}_{\underline{L}}
{\cal M}^{\underline{LK}}
{\cal O}_{\underline{K}}{}^{\underline{N}}~~~.
\nn
\eea

The O(n,n) transformation is recognized as the 
coordinate transformation in the doubled space.
An group element  $\underline{g}(Z^{\underline{M}})$=$
g(Z^M)g'(Z^{M'})=g'(Z^{M'})g(Z^M)$  is transformed 
under the right multiplicative action of the O(n,n) as
\bea
\underline{g}(Z^{\underline{M}})&\to&\check{\underline{g}}(\check{Z}^{\underline{M}})
=\underline{g}({Z}^{\underline{M}})\check{\cal O}
=\underline{g}(Z^{\underline{M}})+\delta {\underline{g}}(Z^{\underline{M}})\label{rightaction}\\
Z^{\underline{M}}&\to&\check{Z}^{\underline{M}}=Z^{\underline{N}}
{\cal O}_{\underline{N}}{}^{\underline{M}}
~~~.\nn
\eea
Under the right multiplicative action which is denoted as
\bea
\Delta=\underline{g}^{-1}\delta \underline{g}=(\check{\cal O}-1)
\label{Dgminus}~~,
\eea
the left-invariant one form $\underline{J}=\underline{g}^{-1}d\underline{g}$
and the right-invariant one form 
${\underline{\tilde{J}}}=d\underline{g}\underline{g}^{-1}$ are
transformed as
\bea
&\delta \underline{J} =d\Delta+[\underline{J},\Delta]
~~,~~
\delta{\underline{\tilde{J}}}=\underline{g}(d\Delta)\underline{g}^{-1}
~~~.&\label{JJO}
\eea
Then the right-invariant one form is inert under the global O(n,n)
transformation, $d\Delta=0~\Rightarrow~\delta\underline{\tilde{J}}=0$.
Hence the symmetry generator  is inert under the global O(n,n) transformation
which rotates the covariant derivative.

\par
\vskip 6mm

\section{Bosonic string action in  doubled space}

We begin with the nondegenerate  Poincar$\acute{\rm e}$ algebra
including both the Lorentz generator and its nondegenerate partner 
\cite{Hatsuda:2001xf,Bonanos:2009wy}.
Then 
it is doubled to construct  chiral affine Poincar\'{e} algebras
\cite{Siegel:2011sy,Polacek:2013nla}
\cite{Hatsuda:2014qqa}-
\cite{Hatsuda:2014aza}.
Concrete expression of the covariant derivatives and the symmetry generators 
for the left and right moving modes are given.
The dimensional reduction constraints  are examined on its consistency, chirality and the O(n,n) transformation.
A gauge invariant bosonic string action is presented, 
and dimensional reduction of the doubled space action
is demonstrated.
\par
\vskip 6mm

\subsection{Doubled chiral  Poincar$\acute{\textbf{e}}$ generators }

Generators of nondegenerate Poincar$\acute{\rm e}$ algebra are given by
 $G_I=( s_{mn},~p_m,~\sigma^{mn})$
 with canonical dimensions $(0,~1,~2)$
  respectively.  The algebra is given as 
\bea
&{\rm dim}~0:& \lbrack s_{mn},s_{lk}]=-i\eta_{[k|[m}s_{n]|l]}\nn\\
&{\rm dim}~1:& \lbrack s_{mn},p_l]=-ip_{[m}\eta_{n]l}~\label{parnonPoi}\nn\\
&{\rm dim}~2:& \lbrack s_{mn},\sigma^{lk}]=-i\delta_{[m}^{[k}\sigma_{n]}{}^{l]}~,~
 \lbrack p_{m},p_{n}]=i\sigma_{mn} 
~.\label{nondegbos}
  \eea
The nondegenerate group metric is
\bea
&&~~~~~~~~~s~~~p~~~\sigma\nn\\
\eta_{IJ}&=&
\begin{array}{c}
s\\p\\\sigma
\end{array}
\left(
\begin{array}{ccc}
&&1\\
&1&\\
1&&\\
\end{array}
\right)~~~.\label{mtrbos}
\eea
The nondegenerate Poincar$\acute{\rm e}$ algebra   \bref{nondegbos}
 is extended to double affine Lie algebras.  
The covariant derivatives and the symmetry generators
 with the constant $b_{IJ}$ solution
in \bref{BnB} are given as follows: 
\begin{itemize}
  \item {Covariant derivatives:
   \bea
{\renewcommand{\arraystretch}{1.6}\begin{array}{lcl}
    {\rm Flat ~left}&:& {\rm Flat ~right}\\    
    ~~ \mathring{\dd}_M=(S_{mn},~P_m,~\Sigma^{mn})&&  
~~\mathring{\dd}_{M'}=(S_{m'n'},~P_{m'},~\Sigma^{m'n'})\\
\left\{\begin{array}{cl}
S_{mn}&=\mathring{\nabla}_S\\
P_m&=\mathring{\nabla}_P+\frac{1}{2}J_1^P\\
\Sigma^{mn}&=\mathring{\nabla}_\Sigma+J_1^S\end{array}\right.
&& 
\left\{\begin{array}{cl}
S_{m'n'}&=\mathring{\nabla}_{S'}\\
P_{m'}&=\mathring{\nabla}_{P'}-\frac{1}{2}J_1^{P'}\\
\Sigma^{m'n'}&=\mathring{\nabla}_{\Sigma'}-J_1^{S'}\end{array}\right.\label{Scovcovb}
 \end{array}} \eea
  }
  \item{Symmetry generators:
\bea
{\renewcommand{\arraystretch}{1.6}
\begin{array}{lcl}
&&{\rm Flat~left}
~~\tilde{\dd}_M=(\tilde{S}_{mn},
\tilde{P}_m,~\tilde{\Sigma}^{mn})
\\
&& \left\{\begin{array}{cl}
\tilde{S}_{mn}&
=\tilde{\nabla}_S-(\tilde{J}_1^\Sigma
+\small{\sum}_{N=S,P} c^N_S\tilde{J}_1^N
)\\
\tilde{P}_m
&=\tilde{\nabla}_P-\frac{1}{2}
(\tilde{J}_1^P+{\small\sum}_{N=S} c^N_P\tilde{J}_1^N
)
\\
\tilde{\Sigma}^{mn}&=\tilde{\nabla}_\Sigma
\end{array}\right.\\
&&{\rm Flat~right}
~~\tilde{\dd}_{M'}=(\tilde{S}_{m'n'},
\tilde{P}_{m'},~\tilde{\Sigma}^{m'n'})
\\
&&
\left\{\begin{array}{cl}
\tilde{S}_{m'n'}
&=\tilde{\nabla}_{S'}+(\tilde{J}_1^{\Sigma'}
+{\small\sum}_{N'=S',P'} c^{N'}_{S'}\tilde{J}_1^{N'}
)\\
\tilde{P}_{m'}
&=\tilde{\nabla}_{P'}+\frac{1}{2}
(\tilde{J}_1^{P'}
+{\small\sum}_{N'=S'} c^{N'}_{P'}\tilde{J}_1^{N'}
)
\\
\tilde{\Sigma}^{m'n'}&=\tilde{\nabla}_{\Sigma'}
\end{array}\right.
\end{array}}
\label{Scovdersymgenb}
\eea }
\end{itemize}
Symmetry generators include 
coefficients $c_M^N$ which are given from \bref{btilb} and \bref{bIJIJ} as
\bea
{\textstyle\sum}_{{N}}^{n_{N}+n_{M}<2} c^{{N}}_{{M}}\tilde{J}_1^{{N}}=J_1^{{N}}B_{{NL}}M_{{M}}{}^{{L}}~~,~~
{\textstyle\sum}_{{N'}}^{n_{N}+n_{M}<2}
 c^{{N}'}_{{M}'}\tilde{J}_1^{{N}'}=-J_1^{{N}'}B_{{N'L'}}M_{{M'}}{}^{{L'}}~~~.
\eea
Nondegeneracy of the algebra requires 
only one chirality for flat left covariant derivative 
in \bref{Scovcovb}
as $\mathring{\nabla}_S$ and 
$\mathring{\nabla}_\Sigma+ J_1^S$,
not both $\mathring{\nabla}_S$ and 
$\mathring{\nabla}_\Sigma\pm J_1^S$.
Therefore \bref{Scovcovb} and \bref{Scovdersymgenb}
 are  unique representation of the affine 
nondegenerate algebras \bref{cov} and \bref{sym}
up to the rescaling of currents.

Rescaled currents with 
parameters $\alpha$ and $\beta$ 
satisfy the following algebra 
modified from \bref{cov}
\bea
  \lbrack \mathring{\dd}_{{I}}(1), 
\mathring{\dd}_{{J}}(2)]=-if_{{IJ}}{}^{{K}}
\mathring{\dd}_{{K}}\delta(2-1)-i\alpha\eta_{{IJ}}
\partial_\sigma\delta(2-1)\label{covalpha}
\eea
as
 \bea
 &&
 {\renewcommand{\arraystretch}{1.6}
\left\{\begin{array}{cl}
S_{mn}&=\mathring{\nabla}_S\\
P_m&=\beta^{-1/2}\mathring{\nabla}_P+
\frac{1}{2}\alpha\beta^{1/2} J_1^P\\
\Sigma^{mn}&=\beta^{-1}\mathring{\nabla}_\Sigma+
\alpha\beta J_1^S\end{array}\right.
}\label{covSPSgb}~~~,
 \eea
and analogous rescaling for another sector.
The usual notation of the left and right moving modes
 is $P_m=\mathring{\dd}_P+J^P_1$
 for $\alpha=2,\beta=1$, or 
  $P_m=\frac{1}{\sqrt{2}}(\mathring{\dd}_P+J^P_1)$ 
  for $\alpha=1,\beta=2$.

The Virasoro operators for a bosonic string 
in a flat space, ${\cal H}_\tau$ and ${\cal H}_\sigma$
are given by
\bea
{\cal H}_\tau&=&\frac{1}{2}\mathring{\dd}_{\underline{M}}\hat{\eta}^{\underline{MN}}\mathring{\dd}_{\underline{N}}\nn\\
&=&\frac{1}{2}P^mP_m+\frac{1}{2}P^{m'}P_{m'}
+\frac{1}{2}\Sigma^{mn}S_{mn}+\frac{1}{2}\Sigma^{m'n'}S_{m'n'}
~\nn
\\
{\cal H}_\sigma&=&\frac{1}{2}\mathring{\dd}_{\underline{M}}\eta^{\underline{MN}}\mathring{\dd}_{\underline{N}}\nn\\
&=&\frac{1}{2}P^mP_m-\frac{1}{2}P^{m'}P_{m'}
+\frac{1}{2}\Sigma^{mn}S_{mn}-\frac{1}{2}\Sigma^{m'n'}S_{m'n'}
~\label{SHH}\label{Virasoro}~~~.
\eea
They satisfy the Virasoro algebra in \bref{ViraVira}.
The two-dimensional chirality is determined by the Virasoro operators
\bref{chiralZM}.
Then  a group element for the nondegenerate Poincar$\acute{\rm e}$ algebra 
in  \bref{nondegbos} is parameterized as
\bea
&&g=g(Z^{\underline{M}})=g(Z^{M}(\sigma^+))g(Z^{M'}(\sigma^-))\nn\\
&&{\renewcommand{\arraystretch}{1.6}
\left\{
\begin{array}{lcl}
g(Z^M)=e^{\frac{i}{2}v_{mn}\sigma^{mn}}e^{ix^mp_m}e^{\frac{i}{2}u^{mn}s_{mn}}
&,&Z^M=(u^{mn},x^m,v_{mn})\\
g(Z^{M'})=e^{\frac{i}{2}v_{m'n'}\sigma^{m'n'}}e^{ix^{m'}p_{m'}}e^{\frac{i}{2}u^{m'n'}s_{m'n'}}&,&Z^{M'}=(u^{m'n'},x^{m'},v_{m'n'})\\
\end{array}\right.}~~~.
\eea
The  $S_{mn}$ component of the left-invariant current 
is given by $g^{-1}dg=
e^{-iu\cdot s}de^{iu\cdot s}+e^{-iu'\cdot s'}de^{iu'\cdot s'} $
with $u\cdot s =\frac{1}{2}u^{mn}s_{mn}$;
\bea
e^{-iu\cdot s}de^{iu\cdot s}
=\frac{e^{{\rm adj}_u}-1}{{\rm adj}_u}idu~\cdot s
~\equiv ~ \Xi^{-1}_uidu~\cdot s~~~,
~~\Xi_u=\displaystyle\frac{{\rm adj}_u}{e^{{\rm adj}_u}-1}~.
\eea
Details are in the appendix \textbf{A}.

Concrete expression of the left-invariant current $J^{\underline{M}}$ and
the right-invariant current $\tilde{J}^{\underline{M}}$  are obtained as follows:
\begin{itemize}
  \item{Left-invariant currents
  \bea
{\renewcommand{\arraystretch}{1.6}
\begin{array}{lcl}
{\rm Flat~left} ~~ J^{{M}}=(J^S,J^P,J^\Sigma)
&:&{\rm Flat ~right}~~J^{{M}'}=(J^{S'},J^{P'},J^{\Sigma'})\\
\left\{
\begin{array}{rcl}
J^S&=&e^{-u}de^{u}=\Xi^{-1}_u du\\
J^P&=&e^{-u}dx\\
J^\Sigma&=&e^{-u}(dv+\frac{1}{2}[x,dx])e^{u}
\end{array}\right.&&
\left\{
\begin{array}{rcl}
J^{S'}&=&-e^{u'}de^{-u'}\\
J^{P'}&=&e^{u'}dx'\\
J^{\Sigma'}&=&e^{u'}(dv'-\frac{1}{2}[x',dx'])e^{-u'}
\end{array}\right. 
\end{array}}\label{JLR}
\eea }

\item{Right-invariant currents 
  \bea
{\renewcommand{\arraystretch}{1.6}
\begin{array}{lcl}
{\rm Flat~left} ~~ \tilde{J}^{{M}}=(\tilde{J}^S,\tilde{J}^P,\tilde{J}^\Sigma)
&:&{\rm Flat ~right}~~J^{{M}'}=(J^{S'},J^{P'},J^{\Sigma'})\\
\left\{
\begin{array}{rcl}
\tilde{J}^S&=&-e^{u}de^{-u}\\
\tilde{J}^P&=&dx-\tilde{J}^Sx\\
\tilde{J}^\Sigma&=&dv-\frac{1}{2}[x,\tilde{J}^P]
+[v,\tilde{J}^S]
\end{array}\right.&&
\left\{
\begin{array}{rcl}
\tilde{J}^{S'}&=&e^{-u'}de^{u'}\\
\tilde{J}^{P'}&=&dx'+\tilde{J}^{S'}x'\\
\tilde{J}^{\Sigma'}&=&dv'+\frac{1}{2}[x',\tilde{J}^{P'}]
+[v',\tilde{J}^{S'}]
\end{array}\right.
\begin{array}{ccl}
\end{array}
\end{array}}\label{currentsKK}
\eea }
\end{itemize}
Detailed computation is given  in the appendix \textbf{B}.
From the above expression it is 
obvious that there is no difference bewteen $J^P$ and $\tilde{J}^P$
if $u$ is absent.

The covariant derivatives 
$\mathring{\nabla}_{\underline{M}}$ 
and the symmetry generators 
$\tilde{\nabla}_{\underline{M}}$ are obtained as follows.
\begin{itemize}
  \item{Covariant derivatives
\bea
{\renewcommand{\arraystretch}{1.6}
\begin{array}{lcl}
{\rm Flat~left}~~\mathring{\nabla}_{M}=(\mathring{\nabla}_S,
\mathring{\nabla}_P,\mathring{\nabla}_\Sigma)&:&
{\rm Flat ~right}~~\mathring{\nabla}_{M'}
=(\mathring{\nabla}_{S'},\mathring{\nabla}_{P'},\mathring{\nabla}_{\Sigma'})\\
\left\{
\begin{array}{rcl}
\mathring{\nabla}_S&=&\Xi_u\frac{1}{i}\partial_u\\
\mathring{\nabla}_P&=&(\frac{1}{i}\partial_x-\frac{1}{2}x \frac{1}{i}\partial_v )e^u\\
\mathring{\nabla}_\Sigma&=&e^{-u}~\frac{1}{i}\partial_v~e^{u}
\end{array}\right.&&
\left\{
\begin{array}{rcl}
\mathring{\nabla}_{S'}&=&\Xi_{-u'}\frac{1}{i}\partial_{u'}\\
\mathring{\nabla}_{P'}&=&(\frac{1}{i}\partial_{x'}+\frac{1}{2}x' \frac{1}{i}\partial_{v'} )e^{-u'}\\
\mathring{\nabla}_{\Sigma'}&=&e^{u'}~\frac{1}{i}\partial_{v'}~e^{-u'}
\end{array}\right.\end{array}}\label{PoCvDr}
\eea}
\item{Symmetry generators
\bea
{\renewcommand{\arraystretch}{1.6}
\begin{array}{lcl}
{\rm Flat~left}~~\tilde{\nabla}_{M}=(\tilde{\nabla}_S,\tilde{\nabla}_P,\tilde{\nabla}_{\Sigma})
&:&{\rm Flat ~right}~~
\tilde{\nabla}_{M'}=(\tilde{\nabla}_{S'},\tilde{\nabla}_{P'},\tilde{\nabla}_{\Sigma'})
\\
\left\{
\begin{array}{l}
\tilde{\nabla}_S=\Xi_{-u}\frac{1}{i}\partial_u+
[x,\frac{1}{i}\partial_x]+ [v, \frac{1}{i}\partial_v] \\
\tilde{\nabla}_P=\frac{1}{i}\partial_x+\frac{1}{2}x \frac{1}{i}\partial_v \\
\tilde{\nabla}_\Sigma=\frac{1}{i}\partial_v
\end{array}\right.&&
\left\{
\begin{array}{l}
\tilde{\nabla}_{S'}=\Xi_{u'}\frac{1}{i}\partial_{u'}-
[x',\frac{1}{i}\partial_{x'}]- [v', \frac{1}{i}\partial_{v'}] \\
\tilde{\nabla}_{P'}=\frac{1}{i}\partial_{x'}-\frac{1}{2}x' \frac{1}{i}\partial_{v'} \\
\tilde{\nabla}_{\Sigma'}=\frac{1}{i}\partial_{v'}
\end{array}\right.
\end{array}
}\nn\\\label{PoSyge}
\eea}
\end{itemize}
In order to distinguish $\mathring{\nabla}_P$ and $\tilde{\nabla}_P$,
$u$ and $v$ are necessary.

Now let us write down concrete expression of the affine algebra generators, 
\bref{Scovcovb} and \bref{Scovdersymgenb}, 
in terms of \bref{JLR}, \bref{currentsKK}, \bref{PoCvDr} and \bref{PoSyge} as 
the followings:
\begin{itemize}
  \item{Covariant derivatives:~
  \bea 
&&{\rm Flat ~left}~\mathring{\dd}_M=(S_{mn},P_m,\Sigma^{mn})~~~~\nn\\&&
{\renewcommand{\arraystretch}{1.6}\left\{\begin{array}{lcl}
S_{mn}&=&\Xi_u\frac{1}{i}\partial_u\\
P_{m}&=&\left(\beta^{-1/2}(\frac{1}{i}\partial_x-\frac{1}{2}\frac{1}{i}x\partial_v)+
\frac{\alpha}{2}\beta^{1/2}\partial_\sigma x
\right)e^{u}
\\
\Sigma^{mn}&=&e^{-u}\left(
\beta^{-1}\frac{1}{i}\partial_v e^u+\alpha\beta\partial_\sigma e^u
\right)
\end{array}\right.}\label{NRcovl}
\\&&
~{\rm Flat ~right}~\mathring{\nabla}_{M'}=(S_{m'n'},P_{m'},\Sigma^{m'n'})
\nn\\&&
{\renewcommand{\arraystretch}{1.6}\left\{\begin{array}{lcl}
S_{m'n'}&=&\Xi_{u'}\frac{1}{i}\partial_{u'}\\
P_{m'}&=&\left(\beta^{-1/2}(\frac{1}{i}\partial_{x'}+\frac{1}{2}\frac{1}{i}x'\partial_{v'})-
\frac{\alpha}{2}\beta^{1/2}\partial_{\sigma} x'
\right)e^{-u'}
\\
\Sigma^{m'n'}&=&e^{u'}\left(
\beta^{-1}\frac{1}{i}\partial_{v'} e^{-u'}
+\alpha\beta\partial_\sigma e^{-u'}
\right)
\end{array}\right.}\label{NRcovr}
\eea}
  \item{Symmetry generators:~~
  \bea
&&{\rm Flat ~left}~\tilde{\dd}_M=(\tilde{S}_{mn}  \tilde{P}_{m}, \tilde{\Sigma}^{mn}  )
\nn\\
&&{\renewcommand{\arraystretch}{1.6}\left\{\begin{array}{lcl}
\tilde{S}_{mn}&=&
\Xi_{-u}\frac{1}{i}\partial_u
-x_{[m}(\frac{1}{i}\partial_{x})_{n]}
-v_{[m|}{}^l(\frac{1}{i}\partial_{v})_{n]l}
-\alpha\partial_\sigma v\\
\tilde{P}_{m}&=&\beta^{-1/2}(\frac{1}{i}\partial_x+\frac{1}{2}\frac{1}{i}
x\partial_v)
-\frac{\alpha}{2}\beta^{1/2}\partial_\sigma x
\\
\tilde{\Sigma}^{mn}&=&
\beta^{-1}\frac{1}{i}\partial_v
\end{array}\right.\label{SymFlatleft}}\\
&&{\rm Flat ~right}~\tilde{\dd}_{M'}=(\tilde{S}_{m'n'},
  \tilde{P}_{m'}, \tilde{\Sigma}^{m'n'}  )
\nn\\&&
{\renewcommand{\arraystretch}{1.6}\left\{\begin{array}{lcl}
\tilde{S}_{m'n'}&=&
\Xi_{u'}\frac{1}{i}\partial_{u'}
+x_{[m'}(\frac{1}{i}\partial_{x'})_{n']}
+v_{[m'|}{}^{l'}(\frac{1}{i}\partial_{v'})_{n']l'}
+\alpha\partial_\sigma v'\\
\tilde{P}_{m'}&=&\beta^{-1/2}(\frac{1}{i}\partial_{x'}-
\frac{1}{2}\frac{1}{i}
x'\partial_{v'})
+\frac{\alpha}{2}\beta^{1/2}\partial_\sigma x'
\\
\tilde{\Sigma}^{m'n'}&=&
\beta^{-1}\frac{1}{i}\partial_{v'}
\end{array}\right.\label{SymFlatright}}
\eea}
\end{itemize}
 
\par\vskip 6mm
 \subsection{Dimensional reduction constraints}

In this section the procedure of dimensional reduction of 
the doubled space into
the usual space is presented.
The doubled space is defined by 
the covariant derivatives $\mathring{\dd}_{\underline{M}}$
given in \bref{Scovcovb}.
Fields are functions of the doubled coordinates $Z^{\underline{M}}$.
This enlarged space contains auxiliary dimensions which are reduced 
if we impose the following constraints.
\begin{enumerate}
\item{Section condition (strong constraint): 

In the curved space covariant derivative operators are 
multiplied with 
the vielbein superfield $E_{\underline{A}}{}^{\underline{M}}$  
\cite{Polacek:2013nla},
\cite{Hatsuda:2014qqa}-
\cite{Hatsuda:2014aza}
.
It can be orthonormal in the doubled space
\bea
\dd_{\underline{A}}=E_{\underline{A}}{}^{\underline{M}}\mathring{\dd}_{\underline{M}}~~,~~
(E^T)^{\underline{M}}{}_{\underline{A}}\eta^{\underline{AB}}
E_{\underline{B}}{}^{\underline{N}}=\eta^{\underline{MN}}~~~.
\eea
The Virasoro operators \bref{Virasoro} in curved space are given as
\bea
{\cal H}_\tau&=&\frac{1}{2}{\dd}_{\underline{A}}\hat{\eta}^{\underline{AB}}{\dd}_{\underline{B}}\nn
\\
{\cal H}_\sigma&=&\frac{1}{2}{\dd}_{\underline{A}}\eta^{\underline{AB}}{\dd}_{\underline{B}}=\frac{1}{2}\mathring{\dd}_{\underline{M}}\eta^{\underline{MN}}\mathring{\dd}_{\underline{N}}
~\label{SHHc}~~,~
\eea
and they are set to be zero as Virasoro constraints.  
 ${\cal H}_\sigma=0$ is background field 
 independent constraint in the doubled  space.
Section condition is the $\sigma$-diffeomorphism  constraint ${\cal H}_\sigma=0$ in the second quantized level 
on all matrix elements as
\bea
{\cal H}_\sigma=
0
&\Rightarrow&
{\cal H}_\sigma|\Phi\rangle =0~{\rm and}~\langle \Psi| {\cal H}_\sigma |\Phi\rangle =0\nn\\
&&{\rm or}\nn\\
&&
{\partial}_{\underline{M}}
\eta^{\underline{MN}}\partial_{\underline{N}}\Phi(Z^{\underline{M}})=0~{\rm and}~
{\partial}_{\underline{M}}\Psi(Z^{\underline{M}})
\eta^{\underline{MN}}\partial_{\underline{N}}\Phi(Z^{\underline{M}})=
0~.\label{Hamperpb}
\eea
This constraint gives rise to the stringy contribution in the ``new Lie derivative" 
\cite{Siegel:1993xq,Hatsuda:2012uk,Hatsuda:2014qqa}.
}

\item{Isotropy constraints:  

For a coset group G/H, with G=Poincar$\acute{\rm e}$ and H= Lorentz, 
Lorentz coordinates are suppressed by  
the isometry constraints \cite{Hatsuda:2001xf,Polacek:2013nla,Hatsuda:2014qqa} 
\bea
S_{mn}=S_{m'n'}=0~~~.\label{cosetb}
\eea
These isotropy constraints are just particle covariant derivatives 
in \bref{Scovcovb}, $\mathring{\nabla}_S=\mathring{\nabla}_{S'}=0$, for 
a constant  solution $b_{IJ}$ in \bref{BnB}.
It allows gauging  away the local Lorentz parameters. 
}
\item{Dimensional reduction constraints: 

For nondegenerate pairs of generators $(S_{mn},~\Sigma^{mn})$
and $(S_{m'n'},~\Sigma^{m'n'})$, 
$\Sigma^{mn}$ and $\Sigma^{m'n'}$ are 
auxiliary dimensions 
so these dimensions should be reduced. 
But $\Sigma^{mn}=\Sigma^{m'n'}=0$ cannot be imposed  as  first class constraints,
since they do not commute with the isotropy constraints \bref{cosetb}.
Instead symmetry generators can be imposed as first class constraints
\cite{Siegel:1994xr,Polacek:2013nla,Hatsuda:2014qqa,Hatsuda:2014aza}
\bea
\tilde{\Sigma}^{mn}=\tilde{\Sigma}^{m'n'}=0~~~. \label{dimredconSigb}
\eea
These constraints become covariant derivatives \bref{Scovdersymgenb}, 
and 
$\tilde{\nabla}_\Sigma=0=\mathring{\nabla}_{\Sigma}=0$
 from  \bref{PoCvDr} and \bref{PoSyge}  for 
a constant  solution $b_{IJ}$ in \bref{BnB}.
It allows gauging  away $\Sigma$ and $\Sigma'$ parameters. 
}

\item{Left/right mixing dimensional reduction constraint :
 
We also impose further dimensional reduction constraint
which mixes the left and right sectors to reduce the doubled space
into the usual space.
Covariant derivatives $P_m$ and $P_{m'}$ are dynamical degrees of freedom
with the Virasoro constraints,
while one combination of symmetry generators can be used as  a first class constraint
 \cite{Polacek:2013nla}
\bea
\tilde{P}_m+\gamma \tilde{P}_{m'}=0\label{PPb} ~~~.
\eea
A commutation relation between constraints in \bref{PPb} requires 
the dimensional reduction constraints $\tilde{\Sigma}^{mn}=\tilde{\Sigma}^{m'n'}=0$ in 
\bref{dimredconSigb}
\bea
&&\lbrack (\tilde{P}_m+\gamma \tilde{P}_{m'})(1), (\tilde{P}_n+\gamma \tilde{P}_{n'})(2)]
\nn\\
&&~~~~~~~=
i(\tilde{\Sigma}_{mn}+\gamma^2\tilde{\Sigma}_{m'n'})\delta(2-1)
+i(1-\gamma^2)\eta_{mn}\partial_\sigma\delta(2-1)
\eea
where the Schwinger term cancels out only for $\gamma^2=1$.
If the constraint is chosen  as $\tilde{P}_m-\tilde{P}_{m'}=0$ with $\gamma=-1$
, then the sum of the left and right momenta corresponds to 
the total momentum for a usual space
$\tilde{P}_m+\tilde{P}_{m'}$. 
}
\end{enumerate}

Summarizing the above, dimensional reduction constraints 
 for a bosonic string in the double coordinates space
are isotropy constraints \bref{cosetb},
dimensional reduction constraints \bref{dimredconSigb} and
the left/right mixing dimensional reduction constraint \bref{PPb}
in addition to the section constraint \bref{Hamperpb}.

The two-dimensional diffeomorphisms are 
modified with the dimensional reduction constraints as
\bea
\partial_+\Phi&=&i\lbrack \textstyle{\int}d\sigma~
\left({\cal A}+\mu S+\tilde{\mu}\tilde{\Sigma}+
\tilde{\mu}^P(\tilde{P}+\gamma\tilde{P}')\right),\Phi]\nn\\
\partial_-\Phi&=&i\lbrack \textstyle{\int}d\sigma~
\left({\cal A}'+\mu' S'+\tilde{\mu}'\tilde{\Sigma}'+
\tilde{\mu}^P(\tilde{P}+\gamma\tilde{P}')\right),\Phi]~~~.
\eea
The covariant derivatives are still manifestly chiral up to
the trivial Lorentz rotation
\bea
\partial_- \mathring{\dd}_M=0=\partial_+ \mathring{\dd}_{M'}~~~,
\eea
but not the symmetry generators in general.

Now we go back to the T-duality symmetry.
Let us examine the global O(n,n) transformation on 
constraints.
The section condition  ${\cal H}_\sigma=0$
is inert under the global O(n,n).
The dimensional reduction constraints 
$\tilde{\Sigma}^{mn}=\tilde{\Sigma}^{m'n'}=\tilde{P}_m+\gamma\tilde{P}_{m'}=0$ are also inert 
under the global O(n,n) transformation as shown in \bref{JJO}.
The isotropy constraints are transformed under  O(n,n)$\ni{\cal O}_{\underline{M}}{}^{\underline{N}}$; so we must impose
the consistency condition as
\bea
&S_{mn}=S_{m'n'}=0~~,~~\delta \mathring{\dd}_{\underline{S}}={\cal O}_{\underline{S}}{}^{\underline{M}}\mathring{\dd}_{\underline{M}}=0\label{n2d}~~~.&
\eea
The dimensional reduction constraints 
$\tilde{\Sigma}^{mn}=\tilde{\Sigma}^{m'n'}=0$ coincide with 
$\Sigma^{mn}=\Sigma^{m'n'}=0$ in the unitary gauge 
\bref{Scovcovb} and  \bref{Scovdersymgenb}.
Then analogously to the above
\bea
&\Sigma^{mn}=\Sigma^{m'n'}=0~~,~~\delta \mathring{\dd}_{\underline{\Sigma}}={\cal O}_{\underline{\Sigma}}{}^{\underline{M}}\mathring{\dd}_{\underline{M}}=0\label{n2dSg}~~~.&
\eea
From the orthonormal property an O(n,n) matrix must satisfies
\bea
{\cal O}_{\underline{S}~\underline{M}}={\cal O}_{\underline{\Sigma}~\underline{M}}=
{\cal O}_{\underline{M}~\underline{S}}={\cal O}_{\underline{M}~\underline{\Sigma}}=0~~~.
\label{Onndd}
\eea
As a result the survived symmetry is O(d,d), ${\cal O}_{\underline{PP}}$ , which is the T-duality symmetry
in the usual space.

\par\vskip 6mm

\subsection{Bosonic string action}

The Hamiltonian for a  bosonic string in the doubled space 
is given by
\bea
{\cal H}&=&
\lambda_\tau {\cal H}_\tau+\lambda_\sigma {\cal H}_\sigma
+\mu^{mn}S_{mn}+\mu^{m'n'}S_{m'n'}
+\tilde{\mu}_{mn}\tilde{\Sigma}^{mn}+\tilde{\mu}_{m'n'}\tilde{\Sigma}^{m'n'}\nn\\
&&+\tilde{\mu}^m(\tilde{P}_m-\tilde{P}_{m'})~~~\label{bostham2gi}\\
&=&\frac{1}{2}\mathring{\dd}_{\underline{M}}
(\lambda_\tau\hat{\eta}^{\underline{MN}}+\lambda_\sigma{\eta}^{\underline{MN}})
\mathring{\dd}_{\underline{N}}
+\mu^{\underline{M}}\mathring{\dd}_{\underline{M}}+\tilde{\mu}^{\underline{M}}
\tilde{\dd}_{\underline{M}}\nn
\eea
with
\bea
\mu^{\underline{M}}&=&\left(\mu^S,~\mu^{S'} ~,~{\rm others}=0\right)~~.\nn\\
\tilde{\mu}^{\underline{M}}&=&\left(\tilde{\mu}^{\Sigma},~\tilde{\mu}^{\Sigma'},
\tilde{\mu}^{P},~\tilde{\mu}^{P'}=-\tilde{\mu}^{P} ~,~{\rm others}=0\right)~~.
\eea
An action for the bosonic string is given by 
\bea
I=\displaystyle \int d\tau d\sigma~{\cal L}~~,~~
{\cal L}
=\partial_\tau Z^{\underline{M}}\frac{1}{i}\partial_{\underline{M}}
-{\cal H}~=~J_0{}^{\underline{M}}\mathring{\nabla}_{\underline{M}}-{\cal H}
~~~.
\eea
In this section we obtain an gauge invariant action without specifying a solution
for $B$ field.
From \bref{ddJb} and \bref{BnB} we use
\bea
\mathring{\dd}_{M}&=&\mathring{\nabla}_M+J_1^Kb_{KM}=
\mathring{\nabla}_M+\frac{1}{2}J_1^K(\eta_{KM}+B_{KM})
\nn\\
\tilde{\dd}_{M}&=&\tilde{\nabla}_M+\tilde{J}_1^K\tilde{b}_{KM}
=M_M{}^N(\mathring{\dd}_N-J_1^K\eta_{KN})~~\label{symcovMgib}
\eea
and the similar relations for the right sector.
In general 
$M_I{}^J=(L^{-1}R)_I{}^J$ is triangular matrix 
from the relation of canonical dimensions.
The orthonormality of  $M_I{}^J$
leads to the orthonormality of 
$M_P{}^P$  
\bea
&&~~~~~~~~~S~~~P~~~\Sigma\nn\\
M_{I}{}^{J}&=&
\begin{array}{c}
S\\P\\\Sigma
\end{array}
\left(
\begin{array}{ccc}
\ast&\ast&\ast\\
0&\ast&\ast\\
0&0&\ast\\
\end{array}
\right)~~~
~\Rightarrow~M_{P_l}{}^{P_m}M_{P_k}{}^{P_n}\eta_{P_mP_n}=\eta_{P_lP_k}~~
\label{MMeta}
\eea 
where $\ast$'s  denote nonzero elements.
The matrices 
$M_P{}^P$ and $M_{P'}{}^{P'}$ are 
 Lorentz  rotation  matrices, and they are 
functions of only Lorentz parameters with the canonical dimension 0.
The triangular property of $M_I{}^J$ leads to
\bea
\tilde{P}&=&M_P{}^P(P-J_1^P)
+M_P{}^\Sigma
(M_\Sigma{}^\Sigma)^{-1} ~\tilde{\Sigma}\nn\\
\tilde{P}'&=&M_{P'}{}^{P'}(P'+J_1^{P'})
+M_{P'}{}^{\Sigma'}
(M_{\Sigma'}{}^{\Sigma'})^{-1} ~\tilde{\Sigma}'~~~\label{PtilPtil}\\
\mathring{\nabla}_\Sigma&=&
\frac{1}{2}J_1{}^S-\frac{1}{2}J_1{}^KB_{K\Sigma}+(M_\Sigma{}^\Sigma)^{-1}\tilde{\Sigma}~~.\nn
\eea

Using \bref{symcovMgib} and \bref{PtilPtil} the Lagrangian becomes 
sum of the kinetic part ${\cal L}_0$, 
the Wess-Zumino terms ${\cal L}_{WZ}$, 
the boundary of the Wess-Zumino terms ${\cal L}_{WZ;0}$ 
and constraint part ${\cal L}_{\rm const}$:
\bea
&&{\cal L}~=~{\cal L}_{0}+{\cal L}_{WZ}+{\cal L}_{WZ;0}+{\cal L}_{\rm const} \nn\\
&&{\renewcommand{\arraystretch}{1.8}
\left\{\begin{array}{ccl}
{\cal L}_{0}&=&J_0{}^{{P}}P-\frac{\lambda}{2}P^2-\frac{1}{2}J_0{}^PJ_1{}^P
-\tilde{\mu}^PM_P{}^P(P-J_1{}^P)\\
&&+J_0{}^{{P}'}P'-\frac{\lambda'}{2}{P'}^2+\frac{1}{2}J_0{}^{P'}J_1{}^{P'}
+\tilde{\mu}^PM_{P'}{}^{P'}(P'+J_1{}^{P'})\\
{\cal L}_{WZ}&=&-\frac{1}{2}(J_0{}^NJ_1{}^MB_{MN}-J_0{}^{N'}J_1{}^{M'}B_{M'N'})
\\
{\cal L}_{WZ;0}&=&
\frac{1}{2}(J_{[0}{}^\Sigma J_{1]}{}^S
-J_{[0}{}^{\Sigma'} J_{1]}{}^{S'})\\
{\cal L}_{\rm const}&=&
-\hat{\mu}^{S} {S}-\hat{\mu}^{S'} {S}'
-\hat{\mu}^{\Sigma} \tilde{\Sigma}
-\hat{\mu}^{\Sigma'} \tilde{\Sigma}'
\end{array}\right. }
\label{Legendregi}~~
\eea
by redefining Lagrange multipliers as
\bea 
{\renewcommand{\arraystretch}{1.6}
\left\{\begin{array}{lcl}
\lambda=\lambda_\tau+\lambda_\sigma&,&
\lambda'=\lambda_\tau-\lambda_\sigma\\
\hat{\mu}^{S}=\mu^{S}-J_0^S+\frac{\lambda}{2}\Sigma&,&
\hat{\mu}^{S'}=\mu^{S'}-J_0^{S'}+\frac{\lambda'}{2}{\Sigma'}
\\
\hat{\mu}^\Sigma=\tilde{\mu}^\Sigma
-(J_0^{\Sigma}-\tilde{\mu}^PM_P{}^\Sigma)(M_\Sigma{}^\Sigma)^{-1}
&,&
\hat{\mu}^{\Sigma'}=\tilde{\mu}^{\Sigma'}-(
J_0^{\Sigma'}+\tilde{\mu}^{P}M_{P'}{}^{\Sigma'})(M_{\Sigma'}{}^{\Sigma'})^{-1}
\end{array}\right.}~~~.\label{lagmul}
\eea
Variation of the action with respect to $\tilde{\mu}^P$ gives
the left/right mixed dimensional reduction constraint 
\bea
\tilde{P}'-\tilde{P}=
(P'+J_1{}^{P'})-\underline{M}_{P'}{}^P(P-J_1{}^P)=0~~,~~
\underline{M}_{P'}{}^P=(M_{P'}{}^{P'})^{-1}M_{P}{}^P~~.
\eea
$\underline{M}_{P'}{}^{P}$ is a Lorentz rotation matrix 
which relate the left and the right spaces where 
the similar matrix is introduced in \cite{Hassan:1999bv}.
After integrating out both $P$ and $P'$ the 
kinetic part becomes
\bea
{\cal L}_0
&=&\frac{\beta}{2(\lambda+\lambda')}
\left\lbrack  ({J}^{\cal P+}_0)^2
-\alpha^2{\lambda\lambda'}( {J}^{\cal P+}_1)^2
+\alpha(-\lambda+\lambda') {J}^{\cal P+}_0~{J}^{\cal P+}_1\right]+\frac{\alpha\beta}{2}J_{[0}^P~
 J_{1]}^{P'}\underline{M}_{P'}{}^P\nn
\\
 {J}^{\cal P\pm}_i&=&\frac{1}{\sqrt{2}}(J^{P}_i\pm J^{P'}_i\underline{M}_{P'}{}^P) ~~~
\eea
where $\alpha$ and $\beta$  are parameters for normalization 
in \bref{covalpha} and \bref{covSPSgb}.
Two Lagrange multipliers correspond to the worldsheet metric as
\bea
\alpha\lambda=-\frac{h^{01}}{h^{00}}+\frac{1}{\sqrt{-h}h^{00}}~~,~~
\alpha\lambda'=\frac{h^{01}}{h^{00}}+\frac{1}{\sqrt{-h}h^{00}}~~~.
\eea
As a result the gauge invariant action for a bosonic string based on the double
 nondegenerate Poincar$\acute{\rm e}$ algebra is given by
\bea
&&I=\displaystyle\int d\tau d\sigma~ {\cal L}~~,~~{\cal L}={\cal L}_0+{\cal L}_{WZ}+{\cal L}_{WZ;0} \nn\\
&&{\renewcommand{\arraystretch}{1.8}
\left\{\begin{array}{lcl}
{\cal L}_0&=&\frac{\alpha\beta}{2}\left\lbrack
\sqrt{-h}h^{ij} {J}^{\cal P+}_i  {J}^{\cal P+}_j
- \epsilon^{ij}{J}^{\cal P+}_i  {J}^{\cal P-}_j
\right]~~~\label{stactionboson}~~\\
{\cal L}_{WZ}&=&-\frac{\alpha\beta}{2}(J_0{}^NJ_1{}^MB_{MN}-J_0{}^{N'}J_1{}^{M'}B_{M'N'})
\\
{\cal L}_{WZ;0}&=&
\frac{\alpha\beta}{2}(J_{[0}{}^\Sigma J_{1]}{}^S
-J_{[0}{}^{\Sigma'} J_{1]}{}^{S'})
\end{array}\right.}
\eea
If we impose the  simple solution for $B$ field in \bref{BnB},
the Wess-Zumino term ${\cal L}_{WZ}$  is cancelled out  by
the boundary term of the Wess-Zumino term ${\cal L}_{WZ;0}$.
For a general solution of $B$ field ${\cal L}_{WZ}+{\cal L}_{WZ;0}$ is
total derivative terms.

Taking variation of the action \bref{stactionboson} with respect to $Z^{\underline{M}}$
the following first class constraints are derived; 
the
Virasoro constraints in \bref{Virasoro},
isotropy constraints \bref{cosetb},
dimensional reduction constraints \bref{dimredconSigb} and
the left/right mixing dimensional reduction constraint \bref{PPb} 
\bea
{\cal H}_\tau~=~{\cal H}_\sigma~=~
S_{mn}~=~S_{m'n'}~=~
\tilde{\Sigma}^{mn}~=~\tilde{\Sigma}^{m'n'}~=~
\tilde{P}_m-\tilde{P}_{m'}=0\label{1st}~~~.
\eea
The gauge invariance generated by the above first class constraints
is preserved.

\par \vskip 6mm
\subsection{Gauge fixing}

Corresponding to the first class constraints \bref{1st}
we can choose the following gauge fixing conditions.
 For the isotropy constraints \bref{cosetb} 
and the dimensional reduction constraints \bref{dimredconSigb}
\bea
S_{mn}=S_{m'n'}=0
~~,~~
\tilde{\Sigma}^{mn}=\tilde{\Sigma}^{m'n'}=0\nn~~~,
\eea
the simplest gauge is an unitary gauge
\bea
u^{mn}=u^{m'n'}=0 ~~,~~v_{mn}=v_{m'n'}={\rm const.}~~~.
\label{uni}
\eea

Let us introduce two  kinds of coordinates as 
\bea
X^{\rm m}=\frac{1
}{\sqrt{2}}(x^m+x^{m'})&,&
Y_{\rm m}~=~\frac{1
}{\sqrt{2}}(x^m-x^{m'})~~,\label{XYcoord}\\
\frac{\partial}{\partial X^{\rm m}}=\frac{1}{\sqrt{2}}
(\partial_{x^m}+\partial_{x^{m'}})&,&
\frac{\partial}{\partial Y_{\rm m}}=\frac{1}{\sqrt{2}}
(\partial_{x^m}-\partial_{x^{m'}})\nn
\eea
where $X^{\rm m}$ and $Y_{\rm m}$ 
correspond to the usual space coordinate and the dual coordinate
respectively. 
The unitary gauge \bref{uni} allows 
$\tilde{J}^S=\tilde{J}^{S'}=0$,~
$\tilde{\nabla}_P=\mathring{\nabla}_P$,~
$\tilde{\nabla}_{P'}=\mathring{\nabla}_{P'}$,~
 $\tilde{J}^P={J}^P$ and $\tilde{J}^{P'}={J}^{P'}$. 
The momentum operators are rewritten from \bref{SymFlatleft} and \bref{SymFlatright} in the unitary gauge 
in terms of $X^{\rm m}$ and $Y_{\rm m}$ as
\bea
\left\{\begin{array}{ccl}
{P}_m+{P}_{m'}&=&
\displaystyle\frac{\sqrt{2}}{i}\frac{\partial}{\partial X^{\rm m}}
+\frac{1}{\sqrt{2}}\partial_\sigma Y_{\rm m}\nn\\
{P}_m-{P}_{m'}&=&
\displaystyle\frac{\sqrt{2}}{i}\frac{\partial}{\partial Y_{\rm m}}
+\frac{1}{\sqrt{2}}\partial_\sigma X^{\rm m}
\end{array}\right.~~
\left\{\begin{array}{ccl}
\tilde{P}_m+\tilde{P}_{m'}&=&
\displaystyle\frac{\sqrt{2}}{i}\frac{\partial}{\partial X^{\rm m}}-
\frac{1}{\sqrt{2}}\partial_\sigma Y_{\rm m}
\nn\\
\tilde{P}_m-\tilde{P}_{m'}&=&
\displaystyle\frac{\sqrt{2}}{i}\frac{\partial  }{\partial Y_{\rm m} }-
\frac{1}{\sqrt{2}}\partial_\sigma X^{\rm m} 
\end{array}\right.~~.
\eea
They become the usual ``dual coordinate relation"
$\tilde{P}_m\pm \tilde{P}_{m'}\Leftrightarrow \partial_i Y_{\rm m}-\epsilon_{ij}\partial^jX^{\rm m}$
in a flat space.
Contrast to the conventional relation 
$\partial_i Y_{\rm m}-\epsilon_{ij}\partial^jX^{\rm m}=0$ \cite{Duff:1989tf,Tseytlin:1990nb},
we set only one of them to be zero,
 $\tilde{P}-\tilde{P}'=0$. 
This is the left/right mixed dimensional reduction constraint
which allows the following gauge fixing condition on $Y_{\rm m}$.
\bea
\tilde{P}_m-\tilde{P}_{m'}=
\displaystyle\frac{\sqrt{2}}{i}\frac{\partial  }{\partial Y_{\rm m} }-
\frac{1}{\sqrt{2}}\partial_\sigma X^{\rm m}=0~\Rightarrow~\partial_\sigma Y_{\rm m}=0~~~.\label{tilP}
\eea
Then the momentum operators become
\bea
{\renewcommand{\arraystretch}{1.6}
\left\{\begin{array}{ccl}
{P}_m&=&\displaystyle\frac{1}{\sqrt{2}}
(\displaystyle\frac{1}{i}\frac{\partial}{\partial X^{\rm m}}
+\partial_\sigma X^{\rm m})
\\
{P}_{m'}&=&\displaystyle\frac{1}{\sqrt{2}}
(\displaystyle\frac{1}{i}\frac{\partial}{\partial X^{\rm m}}
-\partial_\sigma X^{\rm m})
\end{array}\right.~~
\left\{\begin{array}{ccl}
\tilde{P}_m+\tilde{P}_{m'}&=&
\displaystyle\frac{\sqrt{2}}{i}\frac{\partial}{\partial X^{\rm m}}
\\
\tilde{P}_m-\tilde{P}_{m'}&=&0
\end{array}\right.~~}~~~,
\eea
where $P_m$ and $P_{m'}$ are the left and right moving modes
in the usual space 
and the $\tilde{P}_m+\tilde{P}_{m'}$ is the total momentum
of the space.

The section condition \bref{Hamperpb} becomes simpler 
in the unitary gauge  
\bea
{\cal H}_\sigma
\approx
(\frac{1}{i}\frac{\partial}{\partial X}+\frac{1}{{2}}\partial_\sigma Y)_{\rm m}
(\frac{1}{i}\frac{\partial}{\partial Y}+
\frac{1}{{2}}\partial_\sigma X)^{\rm m}
=\frac{1}{i}\frac{\partial}{\partial X}{}_{\rm m}\partial_\sigma X^{\rm m}
=0~~~\label{HamperpP}
\eea
where $\approx$ uses the local Lorentz constraints (isotropy constraints).
A  solution in the second quantized level of \bref{tilP} is given by 
  $\Psi(X,Y)=e^{\frac{i}{2}\int Y\cdot \partial_\sigma X}
\Phi(X)$.
The section condition reduces to the 
usual $\sigma$ diffeomorphism constraint \bref{HamperpP}
in the simple gauge \bref{tilP} as
 $\displaystyle\frac{\partial}{\partial X}{}_{\rm m}\partial_\sigma X^{\rm m}
\Phi(X)=0$ .


There is another simple gauge $u=u'$ in such a way that
$\underline{M}_{P'}{}^P=1$,
which gives ${J}^{\cal P\pm}_i=J_i^P\pm J_i^{P'}$.
In the unitary gauge $u=u'=0$ they become very simple with  
the usual space coordinates \bref{XYcoord};
 ${J}^{\cal P+}_i=\partial_i X$  and 
${J}^{\cal P-}_i=\partial_i Y$.
The Lagrangian for a bosonic string in an unitary gauge $u=u'=0$ 
 is rewritten as
\bea
 {\cal L}_0&=&dX\wedge\ast dX-  dX\wedge dY\label{XYaction}~~~.
\eea
The second term in \bref{XYaction} including the dual coordinate
is a total derivative in a flat space.
This term gives the first class constraint
$\frac{\partial}{i\partial Y^{\rm m}}-\frac{1}{2}\partial_\sigma X^{\rm m}=0$
 in  \bref{tilP}.
Further simple gauge for the section condition \bref{Hamperpb},
$Y_{\rm m}=0$, the action reduces to the usual bosonic string action.

\par\vskip 6mm


\section{Superstring action in  doubled space}

In this section we construct an gauge invariant action for the type II superstring  
in the doubled space.
At first we  present the chiral affine super-Poincar$\acute{\rm e}$ algebras.
The dimensional reduction constraints are extended for a supersymmetric case.
Then we write down an gauge invariant action without using a specific solution of $B$-field.

\subsection{Doubled chiral super-Poincar$\acute{\textbf{e}}$ generators }

For a superstring we use a nondegenerate super-Poincar$\acute{\rm e}$ algebra
generated by 
 $G_I=( s_{mn},~d_\mu,~p_m,~\omega^\mu~,\sigma^{mn})$
 with canonical dimensions $(0,~\frac{1}{2},~1,~\frac{3}{2},~2)$
  respectively.
  The algebra is given as 
\bea
&{\rm dim}~0:& \lbrack s_{mn},s_{lk}]=-i\eta_{[k|[m}s_{n]|l]}\nn\\
&{\rm dim}~\frac{1}{2}:& \lbrack s_{mn},d_\mu]=-\frac{i}{2}(d\gamma_{mn})_\mu\nn\\
&{\rm dim}~1:& \lbrack s_{mn},p_l]=-ip_{[m}\eta_{n]l}~,~
\{d_\mu,d_\nu\}=2p_m\gamma^m{}_{\mu\nu} \label{parnonPoi2}\label{nondegdlas}\\
&{\rm dim}~\frac{3}{2}:&\lbrack s_{mn},\omega^\mu]=\frac{i}{2}(\gamma_{mn}\omega)^{\mu}~,~\lbrack d_\mu,p_m]=2(\gamma_n\omega)_\mu\nn\\ 
&{\rm dim}~2:& \lbrack s_{mn},\sigma^{lk}]=-i\delta_{[m}^{[k}\sigma_{n]}{}^{l]}~,~
 \{d_\mu,\omega^{\nu}\}=-\frac{i}{4}\sigma^{mn}(\gamma_{mn})^\nu{}_\mu~,~
 \lbrack p_{m},p_{n}]=i\sigma_{mn} 
~.\nn
  \eea
The nondegenerate group metric is
\bea
&&~~~~~~~~~s~~~~d~~~p~~~\omega~~~\sigma\nn\\
\eta_{IJ}&=&
\begin{array}{c}
s\\d\\p\\\omega\\\sigma
\end{array}
\left(
\begin{array}{ccccc}
&&&&1\\
&&&1&\\
&&1&&\\
&-1&&&\\
1&&&&
\end{array}
\right)~~~.\label{mtrs}
\eea

The nondegenerate super-Poincar$\acute{\rm e}$ algebra \bref{nondegdlas} is extended to double affine Lie algebras.  
The covariant derivatives and the symmetry generators \bref{covsymbb} are given as follows: 
\begin{itemize}
  \item {Covariant derivatives:
   \bea
{\renewcommand{\arraystretch}{1.6}\begin{array}{lcl}
    {\rm Flat ~left}&:& {\rm Flat ~right}\\    
    ~~ \mathring{\dd}_M=(S_{mn},~D_\mu,~P_m,~\Omega^\mu,~\Sigma^{mn})&&  
~~\mathring{\dd}_{M'}=(S_{m'n'},~D_{\mu'},~P_{m'},~\Omega^{\mu'},~\Sigma^{m'n'})\\
\left\{\begin{array}{cl}
S_{mn}&=\mathring{\nabla}_S\\
D_{\mu}&=\mathring{\nabla}_D-\frac{1}{4}J_1^{\Omega}\\
P_m&=\mathring{\nabla}_P+\frac{1}{2}J_1^P\\
\Omega^\mu&=\mathring{\nabla}_{\Omega}+\frac{3}{4}J_1^D\\
\Sigma^{mn}&=\mathring{\nabla}_\Sigma+J_1^S\end{array}\right.
&& 
\left\{\begin{array}{cl}
S_{m'n'}&=\mathring{\nabla}_{S'}\\
D_{\mu'}&=\mathring{\nabla}_{D'}+\frac{1}{4}J_1^{\Omega'}\\
P_{m'}&=\mathring{\nabla}_{P'}-\frac{1}{2}J_1^{P'}\\
\Omega^{\mu'}&=\mathring{\nabla}_{\Omega'}-\frac{3}{4}J_1^{D'}\\
\Sigma^{m'n'}&=\mathring{\nabla}_{\Sigma'}-J_1^{S'}\end{array}\right.\label{Scovcov}
 \end{array}} \eea
  }
  \item{Symmetry generators:
\bea
{\renewcommand{\arraystretch}{1.6}
\begin{array}{lcl}
&&{\rm Flat~left}
~~\tilde{\dd}_M=(\tilde{S}_{mn},~\tilde{D}_\mu,~
\tilde{P}_m,~\tilde{\Omega}^{\mu},~\tilde{\Sigma}^{mn})
\\
&& \left\{\begin{array}{cl}
\tilde{S}_{mn}&
=\tilde{\nabla}_S-(\tilde{J}_1^\Sigma
+\small{\sum}_{N=S,D,P,\Omega} c^N_S\tilde{J}_1^N
)\\
\tilde{D}_{\mu}&
=\tilde{\nabla}_D+\frac{3}{4}
(\tilde{J}_1^\Omega+{\small\sum}_{N=S,D,P} c^N_D\tilde{J}_1^N
)
\\
\tilde{P}_m
&=\tilde{\nabla}_P-\frac{1}{2}
(\tilde{J}_1^P+{\small\sum}_{N=S,D} c^N_P\tilde{J}_1^N
)
\\
\tilde{\Omega}^{\mu}
&=\tilde{\nabla}_\Omega-\frac{1}{4}
(\tilde{J}_1^D+ c^S_\Omega\tilde{J}_1^S
)
\\
\tilde{\Sigma}^{mn}&=\tilde{\nabla}_\Sigma
\end{array}\right.\\
&&{\rm Flat~right}
~~\tilde{\dd}_{M'}=(\tilde{S}_{m'n'},~\tilde{D}_{\mu'},~
\tilde{P}_{m'},~\tilde{\Omega}^{\mu'},~\tilde{\Sigma}^{m'n'})
\\
&&
\left\{\begin{array}{cl}
\tilde{S}_{m'n'}
&=\tilde{\nabla}_{S'}+(\tilde{J}_1^{\Sigma'}
+{\small\sum}_{N'=S',D',P',\Omega'} c^{N'}_{S'}\tilde{J}_1^{N'}
)\\
\tilde{D}_{\mu'}
&=\tilde{\nabla}_{D'}-\frac{3}{4}
(\tilde{J}_1^{\Omega'}
+{\small\sum}_{N'=S',D',P'} c^{N'}_{D'}\tilde{J}_1^{N'}
\\
\tilde{P}_{m'}
&=\tilde{\nabla}_{P'}+\frac{1}{2}
(\tilde{J}_1^{P'}
+{\small\sum}_{N'=S',D'} c^{N'}_{P'}\tilde{J}_1^{N'}
)
\\
\tilde{\Omega}^{\mu'}
&=\tilde{\nabla}_{\Omega'}+\frac{1}{4}
(\tilde{J}_1^{D'}
+ c^{S'}_{\Omega'}\tilde{J}_1^{S'}
)
\\
\tilde{\Sigma}^{m'n'}&=\tilde{\nabla}_{\Sigma'}
\end{array}\right.
\end{array}}
\label{Scovdersymgen}
\eea }
\end{itemize}
Coefficients $c_M^N$ in symmetry generators are
determined from $\tilde{J}_1^N\tilde{b}_{NM}$ terms 
in \bref{btilb} and \bref{bIJIJ}
and they are functions of super-coordinates
with canonical dimensions $n_M+n_N-2$.
From the nondegeneracy of the group
 \bref{Scovcov} and \bref{Scovdersymgen}
 are  unique representation of the affine 
nondegenerate algebras \bref{cov} and \bref{sym} with 
the superspace metric \bref{mtrs}
up to the rescaling of currents.
Rescaling  currents with 
parameters $\alpha$ and $\beta$, in such a way that they satisfy 
the algebra \bref{covalpha}, is given as
 \bea
 &&
 {\renewcommand{\arraystretch}{1.6}
\left\{\begin{array}{cl}
S_{mn}&=\mathring{\nabla}_S\\
D_\mu&=\beta^{-1/4}\mathring{\nabla}_D-
\frac{1}{4}\alpha\beta^{3/4}J_1^\Omega\\
P_m&=\beta^{-1/2}\mathring{\nabla}_P+
\frac{1}{2}\alpha\beta^{1/2} J_1^P\\
\Omega^\mu&=\beta^{-3/4}\mathring{\nabla}_\Omega+
\frac{3}{4}\alpha\beta^{1/4}J_1^D\\
\Sigma^{mn}&=\beta^{-1}\mathring{\nabla}_\Sigma+
\alpha\beta J_1^S\end{array}\right.
}\label{covSPSg}~~~.
 \eea
The same rescaling is done for other sectors.

\par\vskip 6mm
\subsection{Superstring action}

In this section the supersymmetric extension of the section 3 is  presented.
 The doubled space is defined by 
 the supercovariant derivatives $\mathring{\dd}_{\underline{M}}$ which are 
given in \bref{Scovcov} and  
 parameterized by doubled super-coordinates $Z^{\underline{M}}$.
This enlarged space contains auxiliary  dimensions which are reduced by a set of
first class constraints as same as the bosonic case:
1. the section condition (strong constraints), 2. isotropy constraints
and 4. Left/right mixing dimensional  constraint are the same as before.
Only the condition 3. dimensional reduction constraints for fermions are added.
\begin{description}
  \item[]{3. Dimensional reduction constraints: 

In order to describe correct physical degrees of freedom
unphysical dimensions introduced by  nondegeneracy of the group
are eliminated.
In order to preserve  isotropy constraints and the $\kappa$-symmetry,
symmetry generator currents are chosen to be constraints to
reduce auxiliary dimensions as\cite{Polacek:2013nla,Hatsuda:2014qqa,Hatsuda:2014aza}
\bea
\tilde{\dd}_{\underline{M}}=0~~{\rm for }~~n_{\underline{M}}>1~~\Leftrightarrow~~
\tilde{\Sigma}^{mn}=\tilde{\Sigma}^{m'n'}=\tilde{\Omega}^\mu=\tilde{\Omega}^{\mu'}=0 \label{dimredconSig}~~~.
\eea
 } 
\end{description}

For superstrings the Virasoro constraints are
extended to the $\kappa$-symmetric Virasoro constraints 
${\cal ABCD}$  constraints \cite{Siegel:1985xj,Hatsuda:2001xf,Hatsuda:2014aza} as
\bea
{\cal A}&=&\frac{1}{2}P_mP^m+\Omega^{\mu} D_\mu+
\frac{1}{2}\Sigma^{mn}S_{mn}=0\label{sHamtau}~\\
{\cal B}^\mu&=&P_m(\gamma^mD)^\mu-iS_{mn}(\gamma^{mn}\Omega)^\mu=0\nn\\
{\cal C}_{\mu\nu}&=&D_\mu D_\nu+\frac{1}{2i}S_{mn}P_l(\gamma^{mnl})_{\mu\nu}=0\nn\\
{\cal D}_{m}&=&D\gamma_m\partial_\sigma D+\frac{4}{i}\Sigma_{mn}S^{nl}P_l=0\nn~~~.
\eea
The same set of constraints for the right sector.
First class constraints for a type II superstring in a flat space 
are ${\cal ABCD}$ constraints \bref{sHamtau},
isotropy constraints \bref{cosetb},
dimensional reduction constraints \bref{dimredconSigb} and
the left/right mixing dimensional reduction constraint \bref{PPb}
\bea
{\cal A}~=~{\cal B}~=~{\cal C}~=~{\cal D}~=~
S_{mn}~=~\tilde{\Sigma}^{mn}~=~\tilde{\Omega}^\mu~=~
\tilde{P}_m-\tilde{P}_{m'}=0\nn~~~
\eea
and the similar constraints for the right sector.

 The Hamiltonian for a type II superstring in T-duality covariant form 
is given by
\bea
{\cal H}&=&
\lambda {\cal A}+\lambda_\mu{\cal B}^\mu+\lambda^{\mu\nu}{\cal C}_{\mu\nu}+\lambda^m{\cal D}_m
+\mu^{mn}S_{mn}+\tilde{\mu}_\mu\tilde{\Omega}^\mu+\tilde{\mu}_{mn}\tilde{\Sigma}^{mn}
\nn\\&&
+\lambda'{\cal A}+\lambda_{\mu'}{\cal B}^{\mu'}
+\lambda^{\mu'\nu'}{\cal C}_{\mu'\nu'}
+\lambda^{m'}{\cal D}_{m'}
+\mu^{m'n'}S_{m'n'}
+\tilde{\mu}_{\mu'}\tilde{\Omega}^{\mu'}+\tilde{\mu}_{m'n'}\tilde{\Sigma}^{m'n'}
\nn\\
&&
+\tilde{\mu}^m(\tilde{P}_m-\tilde{P}_{m'})\label{Hmtnsust}~~~\label{bostham2gi}\\
&=&\frac{1}{2}\mathring{\dd}_{\underline{M}}
(\lambda_\tau\hat{\eta}^{\underline{MN}}+\lambda_\sigma{\eta}^{\underline{MN}}+\lambda\cdot\rho^{\underline{MN}})
\mathring{\dd}_{\underline{N}}
+\mu^{\underline{M}}\mathring{\dd}_{\underline{M}}+\tilde{\mu}^{\underline{M}}
\tilde{\dd}_{\underline{M}}\nn
\eea
with
\bea
\mu^{\underline{M}}&=&\left(\mu^S,~\mu^{S'} ~,~{\rm others}=0\right)~~.\nn\\
\tilde{\mu}^{\underline{M}}&=&\left(\tilde{\mu}^{\Sigma},~\tilde{\mu}^{\Sigma'},\tilde{\mu}^{\Omega},~\tilde{\mu}^{\Omega'},~
\tilde{\mu}^{P},~\tilde{\mu}^{P'}=-\tilde{\mu}^{P} ~,~{\rm others}=0\right)~~.
\eea
The matrices $\rho^{\underline{MN}}$ are nilpotent metrics 
introduced to represent ${\cal BCD}$ constraints\cite{Hatsuda:2014qqa}
\bea
&&(\rho^{{MN}})^\mu\mathring{\dd}_M\mathring{\dd}_N={\cal B}^\mu
,~(\rho^{{MN}})_{\mu\nu}\mathring{\dd}_M\mathring{\dd}_N={\cal C}_{\mu\nu}
,~(\rho^{{MN}})^m\mathring{\dd}_M\mathring{\dd}_N={\cal D}^m
\nn~~~
\eea
and the similar relation for the right sector.
An action for a type II superstring in the doubled space is given by 
\bea
I&=&\int d\tau d\sigma ~{\cal L}~~,~~
{\cal L}=\partial_\tau Z^{\underline{M}}\frac{1}{i}\partial_{\underline{M}}
-{\cal H}~\label{sustaction2}
=J_0^{\underline{M}}\mathring{\nabla}_{\underline{M}}-{\cal H}
\eea
The analogous relations in \bref{symcovMgib} and \bref{MMeta} hold for a supersymmetric case.
From the triangle property of the matrix $M_I{}^J$ 
they are rewritten as
\bea
\tilde{P}&=&M_P{}^P(P-J_1^P)
+M_P{}^\Omega(M^{-1})_\Omega{}^\Omega ~\tilde{\Omega}
+\left((M^{-1})_\Omega{}^\Sigma
+M_P{}^\Sigma
(M^{-1})_\Sigma{}^\Sigma\right) ~\tilde{\Sigma}\nn\\
\tilde{P}'&=&M_{P'}{}^{P'}(P'+J_1^{P'})
+M_{P'}{}^{\Omega'}(M^{-1})_{\Omega'}{}^{\Omega'} ~\tilde{\Omega}'
+\left((M^{-1})_{\Omega'}{}^{\Sigma'} 
+M_{P'}{}^{\Sigma'}
(M_{\Sigma'}{}^{\Sigma'})^{-1}\right) ~\tilde{\Sigma}'~~~\nn\\
\mathring{\nabla}_\Omega&=&
\frac{1}{2}(J_1{}^D-J_1{}^KB_{K\Omega})+
(M^{-1})_\Omega{}^\Omega\tilde{\Omega}+
(M^{-1})_\Omega{}^\Sigma\tilde{\Sigma}~~~\nn\\
\mathring{\nabla}_\Sigma&=&
\frac{1}{2}(J_1{}^S-J_1{}^KB_{K\Sigma})
+(M^{-1})_\Sigma{}^\Sigma~ \tilde{\Sigma}~~~.
\eea
Then the Lagrangian is rewritten as
\bea
&&{\cal L}={\cal L}_{0}+{\cal L}_{WZ}+{\cal L}_{WZ;0}+{\cal L}_{\rm const} \label{SUSTaction}\\
&& {\renewcommand{\arraystretch}{1.8}
\left\{\begin{array}{ccl}
{\cal L}_{0}&=&J_0{}^{{P}}P-\frac{\lambda}{2}P^2-\frac{1}{2}J_0{}^PJ_1{}^P
-\tilde{\mu}^P\left(M_P{}^P(P-J_1{}^P)-M_{P'}{}^{P'}(P'+J_1{}^{P'})\right)
\nn\\
&&+J_0{}^{{P}'}P'-\frac{\lambda'}{2}{P'}^2+\frac{1}{2}J_0{}^{P'}J_1{}^{P'}
+\rho D +\rho'D'\nn\\
{\cal L}_{WZ}&=&-\frac{1}{2}(J_0{}^NJ_1{}^MB_{MN}-J_0{}^{N'}J_1{}^{M'}B_{M'N'})\nn\\
{\cal L}_{WZ;0}&=&
\frac{1}{2}(J_{[0}{}^\Sigma J_{1]}{}^S+J_{[0}{}^\Omega J_{1]}{}^D
-J_{[0}{}^{\Sigma'} J_{1]}{}^{S'}-J_{[0}{}^{\Omega'} J_{1]}{}^{D'})\nn\\
{\cal L}_{\rm const}&=&
\rho\cdot D +\rho'\cdot D'
-\hat{\mu}^{S} {S}-\hat{\mu}^{S'} {S}'
-\hat{\mu}^{\Omega} \tilde{\Omega}
-\hat{\mu}^{\Omega'} \tilde{\Omega}'
-\hat{\mu}^{\Sigma} \tilde{\Sigma}
-\hat{\mu}^{\Sigma'} \tilde{\Sigma}'
\end{array}\right.}
~~
\eea
by suitable redefinition of the Lagrange multipliers
similar to \bref{lagmul}
except $\rho$'s
\bea
{\renewcommand{\arraystretch}{1.6}
\begin{array}{ccl}
\rho&=&J_0{}^D-\lambda \Omega-\lambda_\mu \rho^\mu \slP 
-\lambda^{\mu\nu} \rho_{\mu\nu}D-\lambda_m\rho^m
\partial_\sigma D\\
\rho'&=&J_0{}^{D'}-\lambda' \Omega'-\lambda_{\mu'}\rho^{\mu'} \slP' 
-\lambda^{\mu'\nu'}\rho_{\mu'\nu'}D'-\lambda_{m'}\rho^{m'}
\partial_\sigma D'
\end{array}}
\eea
The first class constraints ${\cal BCD}$ are included in
$\rho\cdot D$ and $\rho'\cdot D'$ terms in ${\cal L}_{\rm const}$.

In order to compare it with the Green-Schwarz action 
we use the second class constraints $D_\mu=D_{\mu'}=0$
instead of first class constraints ${\cal BCD}=0$.
The kinetic term becomes the same as the bosonic one ${\cal L}_0$,
while the Wess-Zumino term includes bilinears in the fermionic currents.
We impose the fermionic second class constraints in addition to the first class 
constraints $D_\mu=D_{\mu'}=0$.
The resultant gauge invariant action for a type II superstring  in the
doubled space  is given by
\bea
&&I=\displaystyle\int d\tau d\sigma~{\cal L}~~,~~
{\cal L}~=~{\cal L}_{0}+{\cal L}_{WZ}+{\cal L}_{WZ;0} \nn\\
&&{\renewcommand{\arraystretch}{1.8}
\left\{\begin{array}{ccl}
{\cal L}_{0}&=&\frac{\alpha\beta}{2}\left\lbrack
\sqrt{-h}h^{ij} {J}^{\cal P+}_i  {J}^{\cal P+}_j
- \epsilon^{ij}{J}^{\cal P+}_i  {J}^{\cal P-}_j
\right]~~~\label{staction}~~\\ 
{\cal L}_{WZ}&=&-\frac{\alpha\beta}{2}(J_0{}^NJ_1{}^MB_{MN}-J_0{}^{N'}J_1{}^{M'}B_{M'N'}
)
\\ 
{\cal L}_{WZ;0}&=&
\frac{\alpha\beta}{2}(J_{[0}{}^\Sigma J_{1]}{}^S+J_{[0}{}^\Omega J_{1]}{}^D
-J_{[0}{}^{\Sigma'} J_{1]}{}^{S'}-J_{[0}{}^{\Omega'} J_{1]}{}^{D'})
\end{array}\right.}
\label{SUSTaction2}~~
\eea
where 
\bea
J_i^{{\cal P}\pm}&=&(J_i^P\pm J_i^{P'}\underline{M}_{P'}{}^P)/\sqrt{2}~~~\nn\\
\underline{M}_{{P}'}{}^{P}&=&(M^{-1})_{P'}{}^{P'}M_P{}^P~~~.
\eea 
The normalization parameter $\alpha\beta$ given 
in \bref{covalpha} and \bref{covSPSgb} is natural to set $\alpha\beta=2$.
The Lagrange multipliers $\rho$ and $\rho'$ are
\bea
\rho=J_0^D-\lambda\Omega~,~~\rho'=J_0^{D'}-\lambda'\Omega'~~~.
\eea

Taking variation of the action \bref{SUSTaction2} with respect to
the super-coordinates $Z^{\underline{M}}$
the following first class constraints are derived; 
the fermionic second class and the $\kappa$-symmetry first class constraints,
Virasoro constraints in \bref{Virasoro}, 
the 
isotropy constraints \bref{cosetb},
dimensional reduction constraints \bref{dimredconSigb} and
the left/right mixing dimensional reduction constraint \bref{PPb} 
\bea
&&{\cal H}_\tau~=~{\cal H}_\sigma~=~D_{\mu}~=~D_{\mu'}~=~
S_{mn}~=~S_{m'n'}~=~0\nn\\
&&\tilde{\Omega}^\mu~=~\tilde{\Omega}^{\mu'}~=~\tilde{\Sigma}^{mn}~=~\tilde{\Sigma}^{m'n'}~=~
\tilde{P}_m-\tilde{P}_{m'}=0\label{1stsuper}.
\eea
Type IIA or IIB is determined by the gamma matrix chiral  projection obtained from 
the algebra between supercharges $\tilde{D}_\mu$ and $\tilde{D}_{\mu'}$ and 
the total Lorentz charge $(\tilde{S}-\tilde{S}')_{\rm mn}$.

Under the global O(n,n) transformation,
$S_{mn}$ and $\Sigma^{mn}$ components of the O(n,n) matrix are treated 
as  same as the bosonic case \bref{Onndd}.
Its fermionic components, $D_\mu$ and $\Omega^\mu$, involve
the Ramond-Ramond  dimensions $\Upsilon_{\mu\nu'}$ and $\digamma^{\mu\nu'}$   introduced in \cite{Hatsuda:2014aza}.
This issue will be discussed in another paper.

In the simple solution for $B$ field in \bref{BnB} 
the Lagrangian becomes 
\bea
{\cal L}&=& 
\sqrt{-h}h^{ij} {J}^{\cal P+}_i  {J}^{\cal P+}_j
+ \epsilon^{ij} (-{J}^{\cal P+}_i  {J}^{\cal P-}_j
+\frac{1}{2}J_{i}{}^\Omega J_{j}{}^D-\frac{1}{2}
J_{i}{}^{\Omega'} J_{j}{}^{D'}) \label{SUSTXY}
\eea 
If further simple gauge $\underline{M}_{P'}{}^P=1$
and the constant $B$ field are used, then
it reduces into the $(p,q)$-brane action
proposed by Sakaguchi 
 \cite{Sakaguchi:1998kk}
which is obtained 
from the central extended superalgebra.
 The manifest  SL(2) S-duality
is proposed into the $(p.q)$-brane action
\cite{Abe:1999ph}.
The manifest S and T-duality action will be unified in the F-theory
\cite{Linch:2015lwa}.
With the gauge fixing condition  
$J_i^{{\cal P}-}=0$ and ignoring the surface term,
it reduces to the usual Green-Schwarz superstring action.

 \par\vskip 6mm

 \section{Conclusions}
 
In this paper we have presented
general construction of chiral affine Lie algebras generated by
the supercovariant derivatives and the symmetry generators
for a type II superstring.
The covariant derivatives and the symmetry generators 
have the general form given in \bref{covsymbb}
where the $B$ field is determined from the relation
\bref{btilb} and \bref{bIJIJ}.
There is a constant solution of the $B$ field \bref{BnB}
where the nondegenerate group metric and the dilatation operator
play essential roles.

The obtained covariant derivatives and symmetry generators 
become chiral by doubling the Lie group.
Chirality is manifest in the doubled space;
each coordinate is a function of only the left or right moving coordinate 
in a string worldsheet as
$Z^M(\sigma^+)$ and $Z^{M'}(\sigma^-)$.
Nondegeneracy of the group gives the unique chiral representations.
The supercovariant derivatives are manifestly chiral 
even after the dimensional reduction into the usual space.

The doubled space is reduced into the usual space
by a set of dimensional reduction constraints.
Auxiliary directions introduced for the nondegeneracy of the group
are  reduced by using symmetry generators,
since symmetry generators commute with covariant derivatives.
So the local geometry governed by the covariant derivatives
is preserved under the dimensional reduction.
Therefore the local geometry of the doubled space with 
manifest T-duality is preserved.

Gauge invariant actions for a bosonic string and a type II superstring 
in the doubled space  are obtained in \bref{stactionboson} and 
\bref{SUSTaction2} respectively.
The resultant actions include the kinetic term, the Wess-Zumino
term and the boundary of the Wess-Zumino term.
There exists the winding mode contribution through 
the term $J^{\cal P+}\wedge J^{\cal P-}$,
which can be gauged away by the constraint.

T-duality transformations on branes in the doubled space  
 and the M-theory  and the F-theory 
extension will be interesting issues.
  
 \par\vskip 6mm

\section*{Acknowledgements}
W.S. thanks  to
 Martin Pol\'{a}{\v{c}}ek for valuable discussions. 
M.H. would like to thank the Simons Center for Geometry and Physics for
hospitality during ``the 
2014 Summer Simons workshop in Mathematics and Physics"
where this work has been developed. 
 The work of M.H. is supported  by Grant-in-Aid for Scientific Research (C)
  No. 24540284 from The Ministry of Education, Culture, Sports, Science and Technology of Japan,
and the work of W.S. is
 supported in part by National Science Foundation Grant No. PHY-1316617.

 \par\vskip 6mm

 \appendix
\section*{Appendix}

\section{Composition formula}

We first recall the composition formula valid for small $t$
\bea
&&e^{tA}e^{B}=e^{B+tC_+},~~~~~C_+~=~\sum_{n=0}b_n([B,-)^nA~=~\sum_{n=0}b_n({\rm adj}_B)^nA,\nn
\\
&&e^{B}e^{tA}=e^{B+tC_-},~~~~~C_-~=~\sum_{n=0}(-)^nb_n({\rm adj}_B)^nA,
\\
&&\sum_{r=0}b_r~t^r~=~\frac{t}{e^t-1},
\sum_{r=even}b_r~t^r~=~\frac{t}{2}\frac{e^t+1}{e^t-1},~~~~~~~
\sum_{r=odd}b_r~t^r~=~\frac{-t}{2},
\\
&&b_0~=~1,~~~b_1~=~-\frac12,~~~b_2~=~\frac1{12},~~~b_3~=~0,~~~b_4~=~
-\frac1{720},...\nn
\eea
where in the first line $([B,-)^nA=({\rm adj}_B)^nA$ means the n times commutation
operations between $B$, as~$[B,[B,...[B,A]...]]$.
Consider a operator relation for small $t$
\be
e^{tA}e^{B}~=~e^{C},~~~~C~=~C_0+tC_1+\frac{t^2}{2}C_2+...
\label{hd}
\ee
At $t=0$ ~~$C_0=B$. By taking $t$ derivative
\be
e^{tA}Ae^B~=~\int^1_0 ds e^{sC}(C_1+tC_2+...)e^{(1-s)C}
\ee
and put $t=0$
\bea
A&=&\int^1_0 ds ~e^{sB}C_1e^{-sB}
~=~\int^1_0 ds (C_1+s[B,C_1]+\frac{s^2}{2!}[B,[B,C_1]]+...)\nn
\\
&=&C_1+\frac12[B,C_1]+\frac{1}{3!}[B,[B,C_1]]+...
~=~\sum_{n=0}\frac{1}{(n+1)!}({\rm adj}_B)^nC_1~~~.
\label{10}
\eea
The solution $C_1$ is found in the form
\be
C_1~=~\sum_{n=0}b_n({\rm adj}_B)^nA.
\ee
Put it back into \bref{10}
\be
A~=~\sum_{n=0}\frac{1}{(n+1)!}({\rm adj}_B)^n
\sum_{k=0}b_k({\rm adj}_B)^k A
~=~\sum_{n=0}\sum^n_{k=0}\frac{b_k}{(n-k+1)!}({\rm adj}_B)^n A
\ee
From here we find
\be
b_0~=~1,~~~ {\rm and}~~~
b_r~=~-\sum^{r-1}_{k=0}\frac{b_k}{(r-k+1)!},~(r=1,2,...).
\ee
The latter is solved recursively.
\be
b_0~=~1,~~~b_1~=~-\frac12,~~~b_2~=~\frac1{12},~~~b_3~=~0,~~~b_4~=~
-\frac1{720},...
\ee

We make a generating function
\bea
G(t)&=&\sum_{r=0}b_r~t^r~=~b_0-\sum_{r=1}t^r~\sum^{r-1}_{k=0}
\frac{b_k}{(r-k+1)!}~=~
1-\sum^{\infty}_{k=0}\sum^{\infty}_{r=k+1}\frac{t^r~b_k}{(r-k+1)!}
\nn\\
&=&1-\sum^{\infty}_{k=0}\sum^{\infty}_{s=0}\frac{t^{s+k+1}~b_k}{(s+2)!}
~=~1-\sum^{\infty}_{s=0}\frac{t^{s+1}}{(s+2)!}~G(t).
\eea
From this we can solve $G(t)$ as
\be
G(t)~=~\frac{t}{e^t-1}~=~\frac{1}{1+\frac{t}{2!}+\frac{t^2}{3!}+...}
\ee
The coefficients $b_n$ is found by
\be
b_n~=~\lim_{t\rightarrow 0}\frac{1}{n!}(\pa_t)^n ~G(t).
\ee
Thus the solution of \bref{hd} is
\be
C_1~=~\sum_{n=0}~b_n({\rm adj}_B)^n~A~=~G({\rm adj}_B)~A~=~\frac{({\rm adj}_B)}{e^{({\rm adj}_B)}-1}~A
\ee
$b_n $ is related to the Bernoulli numbers $B_n$ by
\be
b_n\=\frac{B_n}{n!}.
\ee
More generally the  Bernoulli function $B_n(x)$ is defined by
\be
G(t,x)~=~\frac{t\;e^{tx}}{e^t-1}~=~\sum_{n=0}^\infty\frac{t^n}{n!}B_n(x).
\ee
\par\vskip 6mm

  \section{Nonlinear realization of nondegenerate Poincar\'{e} groups }

The left-invariant one form for a Lorentz group is calculated as follows.
Using an abbreviated notation $u\cdot s =\frac{1}{2}u^{mn}s_{mn}$,
the canonical dimension zero part is given $g^{-1}dg=
e^{-iu\cdot s}de^{iu\cdot s}+e^{-iu'\cdot s'}de^{iu'\cdot s'} $;
\bea
e^{-iu\cdot s}de^{iu\cdot s}&=&
\displaystyle\int_0^1 dt ~
e^{-itu\cdot s}~{i}du\cdot s~
e^{itu\cdot s}\nn\\\nn\\
&=&{i}du\cdot s+\frac{1}{2}\lbrack -iu\cdot s,idu\cdot s]
+\frac{1}{3!}\lbrack  -iu\cdot s \lbrack-iu\cdot s,idu\cdot s]]
+\cdots\nn\\\nn\\
&=&\frac{e^{{\rm adj}_u}-1}{{\rm adj}_u}idu~\cdot s
~\equiv ~ \Xi^{-1}_uidu~\cdot s~~~.
\eea
 $\Xi^{-1}$ is an inverse of  $\Xi$ which is given by 
\bea
&\Xi_u=\displaystyle\frac{{\rm adj}_u}{e^{{\rm adj}_u}-1}~,~~
{\rm adj}_u ~A=\lbrack u,A]
&
\eea
expressed by the Bernoulli number $B_n$ as
\bea
&{\displaystyle\frac{q}{e^q-1}}=
{\small\sum}_{n=0}B_n\displaystyle\frac{q^n}{n!}&\\\nn\\
&B_0=1,~B_1=-\displaystyle\frac{1}{2},~B_2=\frac{1}{6},~B_3=0,~B_4=-\frac{1}{30},~\cdots,
B_{2m+1}=0~(m\neq 0)&\nn\\\nn\\
&{\small\sum}_{n={\rm even}}B_n{\displaystyle\frac{q^n}{n!}
=\frac{q}{2}\frac{e^q+1}{e^q-1}}~,~
{\small\sum}_{n={\rm odd}}B_n\displaystyle\frac{q^n}{n!}=-\frac{q}{2}~.&\nn
\eea

The resultant  left-invariant current and right-invariant currents
 of  doubled nondegenerate Poincar$\acute{\rm e}$ algebras 
are  followings:
\begin{itemize}
  \item{Left-invariant currents
  \bea
{\renewcommand{\arraystretch}{1.6}
\begin{array}{l}
{\rm Flat~left} ~~ J^{{M}}=(J^S{}^{mn},J^P{}^{m},J^\Sigma{}_{mn})\\
\left\{
\begin{array}{rcl}
J^S{}^{mn}&=&(e^{-u}){}^{ml}d(e^{u}){}^{ln}=(\Xi^{-1}_u){}^{ml} du{}^{ln}\\
J^P{}^{m}&=&(e^{-u}){}^{mn}dx{}^{n}\\
J^\Sigma{}_{mn}&=&(e^{-u}){}_{ml}(dv{}_{lk}+\frac{1}{2}x_{[l}dx_{k]})(e^{u})_{kn}
\end{array}\right.\\
{\rm Flat ~right}~~J^{{M}'}=(J^{S'}{}^{m'n'},J^{P'}{}^{m'},J^{\Sigma'}{}_{m'n'})\\
\left\{
\begin{array}{rcl}
J^{S'}{}^{m'n'}&=&-(e^{u'})^{m'l'}d(e^{-u'}){}^{l'n'}\\
J^{P'}{}^{m'}&=&(e^{u'})^{m'n'}dx'{}^{n'}\\
J^{\Sigma'}{}_{m'n'}&=&(e^{u'})^{m'l'}(dv'{}_{l'k'}-\frac{1}{2}x'{}_{[l'},dx'{}_{k']})(e^{-u'})_{k'n'}
\end{array}\right. 
\end{array}}\label{JLRap}
\eea }

\item{Right-invariant currents 
  \bea
{\renewcommand{\arraystretch}{1.6}
\begin{array}{l}
{\rm Flat~left} ~~ 
\tilde{J}^{{M}}=(\tilde{J}^S{}^{mn},\tilde{J}^P{}^{m},\tilde{J}^\Sigma{}_{mn})\\
\left\{
\begin{array}{rcl}
\tilde{J}^S{}^{mn}&=&-(e^{u})^{ml}d(e^{-u})^{ln}\\
\tilde{J}^P{}^m&=&dx^m-\tilde{J}^S{}^{mn}x^n\\
\tilde{J}^\Sigma{}_{mn}&=&dv_{mn}-\frac{1}{2}x_{[m}\tilde{J}^P{}_{n]}
+v_{[m|l}\tilde{J}^S{}_{l|n]}
\end{array}\right.\\
{\rm Flat ~right}~~\tilde{J}^{{M}'}=(\tilde{J}^{S'}{}^{m'n'},\tilde{J}^{P'}{}^{m'},\tilde{J}^{\Sigma'}{}_{m'n'})\\
\left\{
\begin{array}{rcl}
\tilde{J}^{S'}{}^{m'n'}&=&(e^{-u'})^{m'l'}d(e^{u'})^{l'n'}\\
\tilde{J}^{P'}{}^{m'}&=&dx'{}^{m'}+\tilde{J}^{S'}{}^{m'n'}x'{}^{n'}\\
\tilde{J}^{\Sigma'}{}_{m'n'}&=&dv'{}_{m'n'}+\frac{1}{2}x'{}_{[m'|l'}\tilde{J}^{P'}{}_{l'|n']}
+v'{}_{[m'|l'}\tilde{J}^{S'}{}_{l'|n']}
\end{array}\right.
\begin{array}{ccl}
\end{array}
\end{array}}\label{currentsKKap}
\eea }
\end{itemize}

 They satisfy the following 
 Maurer-Cartan equations.
 
\begin{itemize}
  \item { The   Maurer-Cartan for the left-invariant currents
  \bea
 && {\renewcommand{\arraystretch}{1.6} 
 \begin{array}{l}{\rm Flat~left} ~~ \\
  \left\{\begin{array}{l}
dJ^S+\,\, J^S \wedge J^S=0\\
dJ^P+\,\,J^S\wedge  J^P=0 \\
dJ^\Sigma+\,\,J^S\wedge J^\Sigma+\,\,J^\Sigma\wedge  J^S -\,\,J^P\wedge  J^P=0
 \end{array}\right.
 \label{MCLIF}
 \\{\rm Flat~right} ~~ \\
\left\{\begin{array}{l}
 dJ^{S'}-\,\, J^{S'}\wedge  J^{S'}=0\\
 dJ^{P'}-\,\,J^{S'}\wedge  J^{P'}=0 \\
dJ^{\Sigma'}-\,\,J^{S'}\wedge J^{\Sigma'}-\,\,J^{\Sigma'} \wedge J^{S'} +\,J^{P'}\wedge  J^{P'}=0
 \end{array}\right.  \end{array}
 }
\eea}
\item{The Maurer-Cartan equations for the right-invariant currents
\bea
&& {\renewcommand{\arraystretch}{1.6}
 \begin{array}{l}{\rm Flat~left} ~~ \\
\left\{ \begin{array}{l}
d\tilde{J}^S-\,\, \tilde{J}^S\wedge  \tilde{J}^S=0
\\
d\tilde{J}^P-\,\,\tilde{J}^S\wedge  \tilde{J}^P=0\\
d\tilde{J}^\Sigma-\,\tilde{J}^S\wedge \tilde{J}^\Sigma-\,\tilde{J}^\Sigma\wedge  \tilde{J}^S+\,\tilde{J}^P\wedge  \tilde{J}^P=0
 \end{array}\right. \\\label{MCRIF}
{\rm Flat~right} ~~ \\
\left\{
 \begin{array}{l}
d\tilde{J}^{S'}+\,\, \tilde{J}^{S'}\wedge  \tilde{J}^{S'}=0
\\
d\tilde{J}^{P'}+\,\,\tilde{J}^{S'}\wedge  \tilde{J}^{P'}=0\\
d\tilde{J}^{\Sigma'}+\,\tilde{J}^{S'}\wedge \tilde{J}^{\Sigma'}
+\,\tilde{J}^{\Sigma'} \wedge \tilde{J}^{S'}
-\,\tilde{J}^{P'}\wedge  \tilde{J}^{P'}=0
 \end{array}\right.\end{array}}
\eea
where Lorentz indices are contracted with  nearest neighbor indices.
}
\end{itemize}


\end{document}